\PassOptionsToPackage{dvipsnames,table}{xcolor}
\documentclass{elsarticle}

\usepackage[utf8]{inputenc}
\usepackage{xcolor} %
\usepackage{amsmath,amsthm,amssymb}
\usepackage{mathtools}
\usepackage{graphicx}
\graphicspath{{./img/}}
\DeclareGraphicsExtensions{.png,PNG,.pdf,.PDF,.eps,.EPS}
\usepackage{enumerate}
\usepackage{subcaption}
\usepackage[colorlinks,linkcolor=blue,urlcolor=blue,citecolor=blue]{hyperref}
\usepackage[percent]{overpic}

\usepackage{booktabs} %
\usepackage{paralist}%
\usepackage{siunitx}%
\usepackage{bm}
\usepackage{float}
\usepackage[ruled,vlined]{algorithm2e}
\usepackage{cleveref} %

\usepackage[acronym,nogroupskip,nonumberlist,automake]{glossaries}
\newcommand{\mat}[1]{\mathbf{{#1}}}
\renewcommand{\vec}[1]{\bm{{#1}}}

\newcommand{\R}{\mathbb{R}}

\DeclareMathOperator{\threshold}{threshold}
\DeclareMathOperator{\ess}{ESS}

\newcommand{\modelupdate}{\mat{F}} %
\newcommand{\state}{\vec{u}} %
\newcommand{\statetruth}{\state^{\textrm{truth}}}
\newcommand{\stateens}{\state^{\textrm{ens}}}
\newcommand{\modelerror}{\vec{\omega}} %
\newcommand{\modelerrorcovariance}{\mat{Q}}
\newcommand{\modeldimension}{M} %

\newcommand{\data}{\vec{y}} %
\newcommand{\dataerror}{\vec{\eta}} %
\newcommand{\dataerrorcovariance}{\mat{R}}
\newcommand{\dataoperator}{\mat{H}}
\newcommand{\datadimension}{D} %

\newcommand{\statereduction}{\mat{V}} %

\newcommand{\rdx}[1]{ {#1}^{q} } %

\newcommand{\reducedmodeldimension}{\rdx{\modeldimension}} %

\newcommand{\reducedmodelerror}{\rdx{\modelerror}} %
\newcommand{\reducedmodelupdate}{\rdx{\modelupdate}} %
\newcommand{\reducedstate}{\rdx{\state}} %
\newcommand{\reducedmodelerrorcovariance}{\rdx{\modelerrorcovariance}}

\newcommand{\reduceddata}{\rdx{\data}} %
\newcommand{\reduceddatadimension}{\rdx{\datadimension}} %
\newcommand{\reduceddataoperator}{\rdx{\dataoperator}}
\newcommand{\reduceddataerror}{\rdx{\dataerror}} %
\newcommand{\reduceddataerrorcovariance}{\rdx{\dataerrorcovariance}}

\newcommand{\datareduction}{\mat{U}} %
\newcommand{\projection}{\mat{\Pi}}

\newcommand{\timeindex}{t} %
\newcommand{\timestepobs}{\tau_{D}}

\newcommand{\timeswitch}{T_F}
\newcommand{\finaltime}{T}
\newcommand{\particleindex}{\ell} %
\newcommand{\totalparticles}{L}
\newcommand{\weight}{w}

\newcommand{\innovation}{\vec{\mathcal{I}}} %

\newcommand{\leftsingularvector}{\vec{\phi}}
\newcommand{\rightsingularvector}{\vec{\psi}}

\newcommand{\singularvalue}{\sigma}

\newcommand{\snapshotmatrix}{\mat{X}}
\newcommand{\probability}{P}

\newcommand{\pinv}{\dagger} %
\newcommand{\id}{\mat{I}} %

\newcommand{\downsamp}[1]{\id_{\downarrow {#1}}}

\DeclarePairedDelimiter{\norm}{\Vert}{\Vert}
\DeclarePairedDelimiter{\abs}{\vert}{\vert}
\DeclarePairedDelimiter{\floor}{\lfloor}{\rfloor}

\DeclareMathOperator{\tol}{tol}

\DeclareMathOperator{\lspan}{span}

\DeclareMathOperator*{\argmin}{arg\,min}

\DeclareMathOperator{\rmse}{RMSE}

\theoremstyle{definition}

\newcommand{\windowsize}{\tau} %
\newcommand{\windowshift}{\tau_{s}} %

\usepackage{comment}

\usepackage{soul} %

\newacronym{DA}{DA}{Data Assimilation}
\newacronym{DMD}{DMD}{Dynamic Mode Decomposition}
\newacronym{POD}{POD}{Proper Orthogonal Decomposition}
\newacronym{SWPOD}{SW-POD}{Sliding Window POD}
\newacronym{SVD}{SVD}{Singular Value Decomposition}
\newacronym{RMSE}{RMSE}{Root Mean Squared Error}
\newacronym{PF}{PF}{Particle Filter}
\newacronym{ProjPF}{Proj-PF}{Projected Particle Filter}

\newacronym{OP-PF}{OP-PF}{Optimal Proposal Particle Filter}
\newacronym{ProjOPPF}{Proj-OP-PF}{Projected Optimal Proposal Particle Filter}
\newacronym{AUS}{AUS}{Assimilation in the Unstable Subspace}
\newacronym{LV}{LV}{Lyapunov Vectors}
\newacronym{SWE}{SWE}{Shallow Water Equations}
\newacronym{ESS}{ESS}{Effective Sample Size}
\newacronym{RESAMP}{RES\%}{Resampling Percentage}
\newacronym{XC}{XC}{Pattern Correlation Coefficient}
\newacronym{L96}{L96}{Lorenz'96 model}
\newacronym{EnKF}{EnKF}{Ensemble Kalman Filter}
\newacronym{KF}{KF}{Kalman Filter}
\newacronym{EKF}{EKF}{Extended Kalman Filter}
\newacronym{PDF}{PDF}{Probability Density Function}
\newacronym{ROM}{ROM}{Reduced Order Models}
\newacronym{4D-Var}{4D-Var}{Four-dimensional variational data assimilation}
\newacronym{I-Proj}{I-Proj}{Proj-OP-PF with the identity projection}
\newacronym{NOSW}{NON-SWPOD}{``Classical'' POD, i.e., without employing the sliding window}

\newglossaryentry{PDE}
{
  type=\acronymtype,
  name={PDE},
  description={partial differential equation},
  first={\glsentrydesc{PDE} (\glsentrytext{PDE})},
  plural={PDEs},
  descriptionplural={partial differential equations},
  firstplural={\glsentrydescplural{PDE} (\glsentryplural{PDE})}
}
\newglossaryentry{ODE}
{
  type=\acronymtype,
  name={ODE},
  description={ordinary differential equation},
  first={\glsentrydesc{ODE} (\glsentrytext{ODE})},
  plural={ODEs},
  descriptionplural={ordinary differential equations},
  firstplural={\glsentrydescplural{ODE} (\glsentryplural{ODE})}
}

\makeglossaries
\usepackage{tikz}
\usepackage{pgfplots,pgfplotstable}
\usetikzlibrary{arrows.meta, positioning, quotes}
\usetikzlibrary{arrows.meta, positioning}

\journal{arXiv} %
\begin{document}

\title{Projected Data Assimilation using Sliding Window Proper Orthogonal Decomposition}

\author[focal1]{Aishah Albarakati}
\ead{aaalbarakati1@uj.edu.sa}
\affiliation[focal1]{organization={Department of Mathematics, University of Jeddah},%
city={Jeddah},
postcode={23218}, 
country={Saudi Arabia}}

\author[focal2]{Marko Budi\v{s}i\'{c}}
\ead{mbudisic@clarkson.edu}
\affiliation[focal2]{organization={Department of Mathematics, Clarkson University},%
city={Potsdam},
postcode={13676}, 
state={NY},
country={USA}}

\author[focal3]{Erik S.
Van Vleck}
\ead{erikvv@ku.edu}
\affiliation[focal3]{organization={Department of Mathematics, University of Kansas},%
  city={Lawrence},
  postcode={66045}, 
  state={KS},
  country={USA}}
  
\begin{abstract}
Prediction of the state evolution of complex high-dimensional nonlinear systems is challenging due to the nonlinear sensitivity of the evolution to small inaccuracies in the model.
\gls{DA} techniques improve state estimates by combining model simulations with real-time data.
Few \gls{DA} techniques can simultaneously handle nonlinear evolution, non-Gaussian uncertainty, and the high dimension of the state.
We recently proposed addressing these challenges using a POD technique that projects the physical and data models into a reduced-dimensional subspace.
\gls{POD} is a tool to extract spatiotemporal patterns (modes) that dominate the observed data.
We combined the \gls{POD}-based projection operator, computed in an offline fashion, with a \gls{DA} scheme that models non-Gaussian uncertainty in lower dimensional subspace.
If the model parameters change significantly during time evolution, the offline computation of the projection operators ceases to be useful.
We address this challenge using a \gls{SWPOD}, which recomputes the projection operator based on a sliding subset of snapshots from the entire evolution.
The physical model projection is updated dynamically in terms of modes and number of modes, and the data model projection is also chosen to promote a sparse approximation.
We test the efficacy of this technique on a modified \gls{L96} with a time-varying forcing and compare it with the time-invariant offline projected algorithm.
In particular, dynamically determined physical and data model projections decrease the \gls{RMSE} and the resampling rate.
\end{abstract}

\begin{keyword}
data assimilation, particle filters, order reduction, proper orthogonal decomposition, Lorenz'96 model.
\end{keyword}

\maketitle

\tableofcontents{%
  \scriptsize%
  \sffamily%
}
\glsreset{DA}
\section{Introduction}
Significant challenges in developing \gls{DA} techniques include nonlinearity, high-dimensional physical and data models, and non-Gaussian posterior distributions.
Few \gls{DA} techniques are capable of simultaneously addressing these problems. In this manuscript, we focus on the \gls{PF} class of algorithms.
Despite particle filters' remarkable ability to handle nonlinearities and non-Gaussian distributions, they suffer from several associated issues.
The \gls{PF} often performs poorly with high-dimensional problems due to what is known as `filter degeneracy' when one of the particle weights approaches one, and all others approach zero, resulting in the posterior distribution being approximately approximated by a single particle.
Filter degeneracy is linked to a phenomenon known as the curse of dimensionality, which affects all sampling algorithms whose efficiency quickly decreases with the increasing dimension of the state space~\cite{Surace2019}.
High-dimensional spaces do not allow adequate resampling to prevent degeneracy, and the number of particles must be extremely large to be considered a good estimate of the posterior~\cite{SnyderEtAl08}, which reduces the computing efficiency of the algorithm.
The\glsreset{OP-PF} \gls{OP-PF} methods \cite{Snyder2011,Snyder15}, have been developed to reduce the sample size necessary to counter particle degeneracy in high dimensions.
Nevertheless, studies analyzing the performance of the \gls{OP-PF} show that the ensemble size required to generate the optimal proposal grows exponentially of the variance of the log-likelihoods of the observations based on the previous state; otherwise, it suffers from filter degeneracy \cite{SnyderEtAl08,Snyder2011}.
Various approaches have been developed to reduce the dimension of the physical state model \cite{wang_proper_2020}, data model \cite{MVV20} or both physical and data models \cite{albarakati2021model} to prevent degeneracy.

Our motivation is based on the development of recent techniques, \gls{ProjOPPF}~\cite{albarakati2021model}, that combined the \glsreset{POD}\gls{POD}-based projection operator, computed in an offline fashion, with the \gls{OP-PF}.
The \gls{POD} can be used to determine the dominant energy modes of medium to high-dimensional models and exploit a possible low-dimensional structure of the model space by reducing it to its corresponding subspace for use in the nonlinear filtering problem \cite{wang_proper_2020}.
Assimilation of projected data involves using the projection operator based on reduced-order models and data assimilation techniques.
In order to calculate the projection operator, we produce a single time-invariant model used throughout the simulation.
However, the offline computation of the projection operators is no longer useful if the model parameters change or the simulation across the periods of the regime change significantly over time.

In this paper, we extend~\cite{albarakati2021model} by employing \glsreset{SWPOD}\gls{SWPOD} to compute projection operators tailored to the moving window of data.
This method can be applied to the analysis of data where specific events occur or change over a relatively short period \cite{quarteroni2014reduced}.
The \gls{SWPOD} has been used to extract the linear and nonlinear modes \cite{clement2014sliding} and also used in data assimilation with machine learning to improve the efficiency of high-dimensional \gls{4D-Var} in \cite{maulik2021efficient}.
The \gls{SWPOD} works by splitting the time domain into subdomains (windows) so that short-lived events or transients affect only a subset of all projections employed.
Furthermore, recomputing of projections allows active tuning of parameters of the projection, e.g., the order of the \gls{ROM}, depending on the quality of the model reduction. For a recent comprehensive review of \gls{ROM} techniques based upon see \cite{SIREV2022} and see also \cite{ZRDC2019, DMM2018, AK2018} for specific techniques that are applicable in the framework developed here.

We employ \gls{SWPOD} to enhance the \gls{OP-PF} assimilation by determining the model and data projections in an offline and online fashion to dynamically determine optimal modes and dimensions and promote sparsity in the case of data-based projections.
The resulting methods adapt to transients in the data to produce effective assimilation with \gls{ProjOPPF} for a nonlinear, high dimensional \gls{L96} model with variable forcing.

In this paper, we present the background about data assimilation techniques in \cref{sec:background_sec}.
We formulate the projected data assimilation using abstract orthogonal projections in \cref{sec:dimension-reduction} and their use in the context of projected optimal proposal particle filters \gls{ProjOPPF} in \cref{sec:ProjOPPF}.
\Cref{sec:tech-model-reduction}, presents the basics of the \gls{POD} and training of mode using \glsreset{SWPOD}\gls{SWPOD}.
The offline fixed dimension scheme is presented in \cref{sec:adaptive-selection-of-parameters}, the offline adaptivity in \cref{sec:offline-adapt} and the online adaptivity in \cref{sec:online-adapt}.
\Cref{sec:results}, contains the numerical results obtained using the \gls{SWPOD} algorithms developed to take advantage of these projected physical and data models from a sliding subset of snapshots.
The methods are applied to the nonlinear \gls{L96} with time-varying forcing in \cref{sec:l96-model}.
The numerical results in \cref{sec:EXP-results} show the efficacy of the \gls{SWPOD} technique compared to the primary \gls{POD} method.
\section{Background on Particle Filters for Reduced-Order Models}
\label{sec:background_sec}
In studying large-scale geophysical systems such as climate, ocean, and atmosphere, \glsreset{DA}\gls{DA} is commonly used to provide accurate estimates of states.
\gls{DA} calculates estimates of the state \(\state_{\timeindex}\) of some physical system at a given time \(\timeindex\) by combining \emph{observations} \(\data_{\timeindex}\) with a dynamical physical model in an optimal way.
A physical state and data models are used to formulate the data assimilation problem.
Consider a state vector \(\state_{\timeindex}\in \mathbb{R}^{\modeldimension}\) that evolves according to the discrete-time stochastic model:
\begin{equation}
\state_{\timeindex}= \modelupdate\left(\state_{\timeindex-1 }\right)+\modelerror_{\timeindex}, \quad\quad \text{(Physical model)}\label{eq:physical-model}
\end{equation}
where \(\modelupdate:\R^\modeldimension\rightarrow \R^\modeldimension\) is a deterministic function of the state \(\state_{\timeindex-1}\) and \(\modelerror_{\timeindex}\) is the state noise vector that models the uncaptured physics, or randomness inherent in the physical process.
We assume \(\modelerror_{\timeindex}\) is Gaussian  \(\modelerror_{\timeindex} \sim \mathcal{N}(0,\modelerrorcovariance_{\timeindex})\), with covariance matrix \(\modelerrorcovariance_{\timeindex}\).
The data model relates the observation vector \(\data_{\timeindex}\in \mathbb{R}^\datadimension\) to the state by
\begin{equation}
\data_{\timeindex} = \dataoperator\state_{\timeindex} + \dataerror_{\timeindex}. \quad\quad \text{(Data model)}\label{eq:data}
\end{equation}
We will assume a linear observation operator \(\dataoperator :\R^\modeldimension\rightarrow \R^\datadimension,\, \datadimension \leq \modeldimension\) and that the observation noise vector \(\dataerror_{\timeindex} \sim  \mathcal{N}\left(0,\dataerrorcovariance_{\timeindex}\right)\), is normally distributed with a covariance matrix \(\dataerrorcovariance_{\timeindex}\).

The initial state \(\state_{0}\), and the noise vectors \(\modelerror_{\timeindex}\) and \(\dataerror_{\timeindex}\) at each step are all assumed to be independent of each other.

The \eqref{eq:data} implies that the data \(\data_{\timeindex}\) is generated from the true (unknown) state by \(\data_{\timeindex} = \dataoperator\state^\text{truth}_{\timeindex} + \dataerror_{\timeindex}\), and then there is a framework that converts the estimates of the state \(\state_{\timeindex}\) into `data space' via  \(\dataoperator\state_{\timeindex}\).

Since we are interested in \gls{DA} problems in which the model \eqref{eq:physical-model} is nonlinear, particle filters are among the most common \gls{DA} methods that work with nonlinear systems, as they are capable of reproducing the true target state in the case of large numbers of particles~\cite{KodyBook}.
\glsreset{PF}\glsreset{OP-PF}
The \glsreset{PF}\gls{PF} approach~\cite{Vetra2018, Leeuwen2015} is based on Monte Carlo sampling of the uncertainty distribution.
\gls{PF} uses a set of state vectors (called \emph{particles}) \(\state^{\particleindex}_{\timeindex - 1}\) and associated non-negative \emph{weights} \(\weight_{\timeindex-1}^{\particleindex}\), with \(\sum_{\particleindex} \weight_{\timeindex-1}^{\particleindex} = 1\), as a discrete probability model for the uncertainty in the state estimate.

The density of the uncertainty distribution model for the state at time \(\timeindex\) is
\begin{equation}
    \label{eq:pfDist}
    \probability_{{\timeindex }}(\state) = \sum_{\particleindex=1}^\totalparticles w_{\timeindex}^{\ell} \delta\left(\state-\state^{\particleindex}_{\timeindex}\right),
\end{equation}
\begin{equation}
   \stateens_{\timeindex} \coloneqq \int \state \probability_{{\timeindex }}(\state) d\state
    = \sum_{\particleindex=1}^\totalparticles w_{\timeindex}^{\ell} \state^{\particleindex}_{\timeindex}.
    \label{eq:ensemble-estimate}
\end{equation}
Though the \gls{PF} algorithm is suitable for assimilation of nonlinear systems, it suffers from several associated issues, including \emph{filter degeneracy}.

The \gls{PF} models the distribution of uncertainty by running several copies (particles) of the model in parallel, with a weight coefficient that is progressively increased if the particle agrees with incoming measurements.
The ensemble estimate  \eqref{eq:ensemble-estimate} is computed as a weighted average of particle states.
If all except one of the particles acquire zero-weight, known as filter degeneracy, \gls{PF} cannot continue effectively capturing the uncertainty spread.
At that moment, the particles are \emph{resampled} to restore their weights, although this affects the continuity of data assimilation and potentially leads to a decrease in the quality of the ensemble estimate.

To detect degeneracy, one monitors the \gls{ESS}, for non-negative weights such that
${\sum_{\particleindex=1}^\totalparticles\left(\weight^{\particleindex}\right)}=1,$
\begin{equation}
    \operatorname{ESS} \coloneqq 
    \left[{\sum_{\particleindex=1}^\totalparticles \left(\weight^{\particleindex}\right)^2}\right]^{-1}
    \label{eq:ESS1}
    \end{equation}
    which always satisfies \(1 \leq \operatorname{ESS} \leq \totalparticles\).
    Ideally, \(\operatorname{ESS} = \totalparticles\); if \gls{ESS} drops  below a chosen threshold, the particles are resampled using one of the standard algorithms (see for example~\cite[\S3.3.2]{HandbookDA}).
    
Filter degeneracy can be mitigated by increasing the number of particles~\cite{Surace2019}, although the required number scales exponentially with the dimension of the space~\cite{SnyderEtAl08,Snyder2011,Snyder15}.
Alternatively, our previous work \cite{MVV20,albarakati2021model} has shown that employing model reduction techniques to the physical model or the data model is efficacious in preventing degeneracy.

\subsection{Reduction of state and data models via orthogonal projections}
\label{sec:dimension-reduction}
Both physical model \eqref{eq:physical-model} and data model \eqref{eq:data} can be projected to reduce dimension.
Starting with the reduction of the dimension of the state \(\state_{\timeindex} \in \mathbb{R}^\modeldimension\).
Consider a matrix \(\statereduction_{\timeindex} \in \mathbb{R}^{\modeldimension \times \reducedmodeldimension}\) whose columns form an orthonormal basis (\(\statereduction_{\timeindex}^{\top}\statereduction_{\timeindex}\equiv\id\)) for a time-dependent subspace on which we are projecting the models.
The map  \(\projection_{\timeindex}^\modeldimension : \mathbb{R}^{\modeldimension} \to \mathbb{R}^{\modeldimension}\), \( \projection_{\timeindex}^\modeldimension = \statereduction_{\timeindex}\statereduction_{\timeindex}^{\top}\) is the orthogonal projection onto the \(\lspan \statereduction_{\timeindex}\), which is interpreted as a composition of the \emph{reduction} and \emph{reconstruction} maps.

The \emph{reduction} of the state vector \(\statereduction_{\timeindex}^{\top}:\mathbb{R}^{\modeldimension} \to \mathbb{R}^{\reducedmodeldimension}\) creates a vector of inner products between the input and the orthonormal basis of the target subspace
\begin{equation}
  \label{eq:projection}
  \reducedstate = \statereduction_{\timeindex}^{\top} \state, \quad \reducedstate \in \mathbb{R}^{\reducedmodeldimension}.
\end{equation}
The \emph{reconstruction} \(\statereduction_{\timeindex}:\mathbb{R}^{\reducedmodeldimension} \to \mathbb{R}^{\modeldimension}\) generates the reconstructed state \(\state^{r}\) as a linear combination of the basis with the coefficients taken from the input vector \(\reducedstate\)
\begin{equation}
\state^{r} \coloneqq \statereduction_{\timeindex} \reducedstate.
\label{eq:reconstruction}
\end{equation}
The output \(\state^{r}\) is an element of the full state space \(\mathbb{R}^{\modeldimension}\), restricted to the \(\lspan \statereduction_{\timeindex}\).

Computing the reduction of the reconstruction, recovers the reduced states \( \statereduction_{\timeindex}^{\top}(\statereduction_{\timeindex} \reducedstate_{\timeindex}) = \reducedstate_{\timeindex},\) due to orthogonality (\(\statereduction_{\timeindex}^{\top}\statereduction_{\timeindex}=\id\)).

Computing the reconstruction of a reduction projects the state onto the subspace, \(\statereduction_{\timeindex}\statereduction_{\timeindex}^{\top} \state_{\timeindex} \in \lspan \statereduction_{\timeindex}\).
Unless the state \(\state_{\timeindex}\) was initially in \(\lspan\statereduction_{\timeindex}\), this map does not reconstruct the input state exactly.

To evolve reduced states \(\reducedstate_{\timeindex}\) using the physical model, we first reconstruct the state to form \(\statereduction_{\timeindex}\reducedstate_{\timeindex}\), apply the evolution map~\eqref{eq:physical-model} to it, and then reduce the output using \(\statereduction_{\timeindex}^{\top}\):
  \begin{align}
  \reducedstate_{\timeindex} &=
            \reducedmodelupdate (\reducedstate_{\timeindex - 1}) + \reducedmodelerror_{\timeindex},\\
            \shortintertext{where}
            \reducedmodelupdate( \reducedstate_{\timeindex-1} ) &= \statereduction_{\timeindex}^{\top} \modelupdate(\statereduction_{\timeindex-1 } \reducedstate_{\timeindex-1}), \quad
            \reducedmodelerror_{\timeindex} \sim  \mathcal{N}(0,\underbrace{\statereduction_{\timeindex}^{\top} \modelerrorcovariance_{\timeindex} \statereduction_{\timeindex}}_{=\reducedmodelerrorcovariance_{\timeindex}}).
  \label{eq:reduced-model-summary}
\end{align}

We can similarly reduce the observation space to another \(\reduceddatadimension\)-dimensional subspace, spanned by columns of \(\datareduction_{\timeindex}\).
The reduced data model \(  \reduceddata_{\timeindex}\) is given by:
\begin{equation}
  \reduceddata_{\timeindex} =\reduceddataoperator \reducedstate_{\timeindex}  +
 \reduceddataerror_{\timeindex},
 \quad
 \reduceddataerror_{\timeindex} \sim \mathcal{N}\left(0,\reduceddataerrorcovariance_{\timeindex}\right),
 \label{eq:reduced-data-summary}
\end{equation}
where the projected observation operator \(\reduceddataoperator\) and the corresponding observation covariance \(\reduceddataerrorcovariance_{\timeindex}\) are given by
\begin{equation}
  \reduceddataoperator =
    \datareduction_{\timeindex}^{\top}\dataoperator^{\pinv}\dataoperator \statereduction_{\timeindex}
  ,\quad
  \reduceddataerrorcovariance_{\timeindex}  =
  \datareduction_{\timeindex}^{\top}\dataoperator^{\pinv}\dataerrorcovariance_{\timeindex}{(\dataoperator^{\pinv})}^{\top} \datareduction_{\timeindex}, \label{eq:physical-model-based-reduction}
\end{equation}
The orthonormal bases \(\statereduction_{\timeindex}\) and \(\datareduction_{\timeindex}\) can be obtained from different dimension reduction techniques, in particular, \gls{POD}, described in more detail in~\cref{sec:pod}.

\subsection{Projected Optimal Proposal Particle Filter (Proj-OP-PF)}
\label{sec:ProjOPPF}
We now summarize the \gls{ProjOPPF} developed in \cite{albarakati2021model} that relies on the described orthogonal projection of model equations.

 Let
\(\reducedstate{}^{\particleindex}_{\timeindex} \coloneqq \statereduction_{\timeindex}^{\top}\state^{\particleindex}_{\timeindex}\) for \(\particleindex=1,\dots,\totalparticles\) denote the \(\particleindex\)th projected particle at time \(\timeindex \) where \(\totalparticles\) is the total number of particles.
Using the projected physical model with the projected data model (in-state model space), together with their corresponding covariance matrices \(\reducedmodelerrorcovariance_{\timeindex} = \statereduction_{\timeindex}^{\top} \modelerrorcovariance_{\timeindex} \statereduction_{\timeindex}\) and \(\reduceddataerrorcovariance_{\timeindex} = \datareduction_{\timeindex}^{\top}\dataoperator^{\pinv}\dataerrorcovariance_{\timeindex}(\dataoperator^{\pinv})^{\top}\datareduction_{\timeindex}\) respectively, with the projected observation operator,
  \(\reduceddataoperator_{\timeindex} = \datareduction_{\timeindex}^{\top}\dataoperator^{\pinv} \dataoperator \statereduction_{\timeindex}\).
  \begin{description}
\item[Projected particle update:]
use the optimal proposal particle update on the projected physical and original data models as:
\begin{align}
    \reducedstate{}^{\particleindex}_{\timeindex} &=  \reducedmodelupdate(\reducedstate{}^{\particleindex}_{\timeindex - 1}) +\reducedmodelerror_{\timeindex}+ \mat{K} \left(\data_{\timeindex} - \dataoperator \statereduction_{\timeindex} \reducedmodelupdate(\reducedstate{}^{\particleindex}_{\timeindex - 1})\right),\,\\
    \shortintertext{where}
    \mat{K}&=\modelerrorcovariance_{p}(\dataoperator \statereduction_{\timeindex})^{\top} \dataerrorcovariance^{-1}_{\timeindex},\,\\
    \modelerrorcovariance^{-1}_{p} &= (\reducedmodelerrorcovariance_{\timeindex})^{-1} +  (\dataoperator \statereduction_{\timeindex})^{\top}  \dataerrorcovariance_{\timeindex}^{-1}  (\dataoperator \statereduction_{\timeindex})
  \end{align}
  \item[Projected weight update:]
  \begin{align}\label{projOPPFweightupdate}
  \weight^{\particleindex}_{\timeindex} \propto& \exp[-\frac{1}{2}(\innovation^{\particleindex}_{\timeindex})^{\top} (\rdx{\mat{Z}}_{\timeindex})^{-1}(\innovation^{\particleindex}_{\timeindex})]\weight_{\timeindex - 1}^{\particleindex},\quad \particleindex=1,\dots,\totalparticles,\,\\
  \shortintertext{where}
  \rdx{\mat{Z}}_{\timeindex}\coloneqq& (\reduceddataoperator_{\timeindex})\reducedmodelerrorcovariance_{\timeindex}(\reduceddataoperator_{\timeindex})^{\top} +\reduceddataerrorcovariance_{\timeindex},\,\\
     \innovation^{\particleindex}_{\timeindex} &= \reduceddata_{\timeindex} - \reduceddataoperator_{\timeindex} \reducedstate{}^{\particleindex}_{\timeindex}.
  \end{align}
  \end{description}
  \glsreset{I-Proj}
  
The unprojected \gls{OP-PF} is recovered by setting projections to identity matrices \(\statereduction_{\timeindex } \equiv \datareduction_{\timeindex } \equiv\id\).
We will refer to the unprojected \gls{OP-PF} as \gls{I-Proj}, when it will be used as the reference model for the projected filter.

\subsection{Resampling scheme}\label{sec:resampling} 
The \gls{ESS} is given by (\ref{eq:ESS1}).
Resampling by an extension of the resampling scheme given in~\cite{MVV20} occurs when the \gls{ESS} falls below a threshold (e.g., \(\ess < \frac{1}{2}\totalparticles\)).
Resampled particles are added by regenerating an unweighted particle ensemble. We then diffuse these particles by adding noise. The noise is generated by sampling gaussian random vectors \(\dataerror \sim \mathcal{N}(\mathbf{0},\omega \id)\), and transforming them as
\begin{equation}\label{resampling}
\statereduction_{\timeindex}^{\top}[\alpha \datareduction_{\timeindex} \datareduction_{\timeindex}^{\top}+(1-\alpha)\id]\dataerror.
\end{equation}
The parameter \( 0 < \alpha < 1\) is the proportion of resampling variance inside the subspace of the reduced data model, \(\lspan \datareduction_{\timeindex}\), and \(\omega\geq 0\) is the (tuneable) total resampling variance.
\section{Time-varying model reduction using sliding-window POD}
\label{sec:tech-model-reduction}
\subsection{Proper Orthogonal Decomposition (POD)}
\label{sec:pod}
\glsreset{POD}\gls{POD} refers to the calculation of orthogonal coordinates for the subspace in which collected data evolve.
It is omnipresent in applied mathematics and is known as principal component analysis (PCA), Karhunen--Lo\'eve decomposition, and empirical orthogonal function (EOF) decomposition in other contexts.
An excellent short review of the main features can be found in \cite[\S 22.4]{Tropea2007}.

 Consider \(\finaltime\) state vectors (called \emph{snapshots}) \(\state_{\timeindex} \in \mathbb{R}^{\modeldimension}\), where \(\timeindex=1,\dots,\finaltime\)
\begin{equation}
  \snapshotmatrix \coloneqq
\begin{bmatrix}
  \state_{1} & \state_{2} & \dots & \state_{\finaltime}
\end{bmatrix}\label{eq:snapshot-matrix}
\end{equation}
A decomposition of the state evolution into a separation of variables ansatz can be written as 
\begin{equation}
 \state_{\timeindex} \approx \sum_{m=1}^{\modeldimension} \leftsingularvector_{m} \singularvalue_{m} \rightsingularvector_{\timeindex ,m}, \label{eq:POD-ansatz}
\end{equation}
where unit-norm vectors \(\leftsingularvector_{m}\), and \(\rightsingularvector_{m}\)  represent the, respectively, ``spatial'' and temporal profiles, associated with the mode \(m\), while \(\singularvalue_{m}\) are the linear combination coefficients.
Although there are many possible separation of variable decompositions, \gls{POD} is characterized by the requirement that both the \(\{\leftsingularvector_{m}\}_{m=1}^\modeldimension\) and \(\{\rightsingularvector_{m}\}_{m=1}^\modeldimension\) should be orthogonal sets.
\gls{POD} can be computed by the \glsreset{SVD}\gls{SVD} of the matrix \(\snapshotmatrix\), writing the factorization in~\eqref{eq:POD-ansatz} as a product of three matrices
\begin{equation}
  \label{eq:svd-of-u}
  \snapshotmatrix =
  \begin{bmatrix}
    \leftsingularvector_{1} &  \leftsingularvector_{2} & \dots
  \end{bmatrix}
  \begin{bsmallmatrix}
    \singularvalue_{1} &  & \\
    & \singularvalue_{2} & & \\
    & & \ddots
  \end{bsmallmatrix}
  \begin{bmatrix}
    \rightsingularvector_{1} &  \rightsingularvector_{2} & \dots
  \end{bmatrix}^{\top}
  ,
\end{equation}
where the entries \(\sigma_{m} \geq 0\) in the diagonal matrix are the singular values, while left- and right-singular vectors, \(\leftsingularvector_{m}\), \(\rightsingularvector_{m}\), are orthonormal bases for the column and row spaces, respectively.
The singular values \(\sigma_{m}\) are commonly ordered in decreasing order, and the \(\sigma_{m}\) and vectors \(\leftsingularvector_{m}\) with a low index are called \emph{dominant}.
In particular, we work with the ``economy'' version of \gls{SVD} that omits the singular values that are equal to zero and their associated singular vectors.
The remaining vectors \(\leftsingularvector_{m}\) and \(\rightsingularvector_{m}\), respectively, span the range and the cokernel of \(\snapshotmatrix\).

To reduce the dimension of \(\snapshotmatrix\), we form the reduction matrix 
 \(\statereduction\) in the following way
\begin{equation}
  \label{eq:POD-projection}
  \mat{\statereduction} =
  \begin{bmatrix}
    \leftsingularvector_{1} & \cdots & \leftsingularvector_{\reducedmodeldimension}
  \end{bmatrix},
\end{equation}
retaining the dominant \(\reducedmodeldimension < \modeldimension\) basis vectors.
According to the Eckart--Young theorem~\cite{Eckart1936}, the projected snapshot matrix
\(\statereduction\statereduction^{\top} \snapshotmatrix \)
is the best approximation of \(\snapshotmatrix\) among all matrices of rank \(\reducedmodeldimension\) as measured by either induced 2-norm or the Frobenius norm; for more see~\cite[\S 2.4]{Golub2013}.

\subsection{Computing model reduction and data reduction matrices using \texorpdfstring{\gls{POD}}{p}}
\label{sec:reduction-using-POD}

The model reduction matrix \(\statereduction\) is computed by applying \gls{POD} to the snapshot matrix \(\snapshotmatrix\), which is equivalent to computing the economy singular value decomposition \(\snapshotmatrix = \mat{\Phi}\mat{\Sigma}\mat{\Psi}\) and retaining first \(\reducedmodeldimension < \modeldimension\) columns, setting \(\statereduction = \mat{\Phi}_{[1:\reducedmodeldimension]}\).

The data reduction matrix \(\datareduction\) can be computed in two non-equivalent ways, as the resulting subspace needs to be in the span of the dominant POD vectors and in the cokernel (input subspace) of the data operator \(H\).
Since projections to POD-dominant vectors and to the cokernel of \(\dataoperator\), \(\projection_{\dataoperator} \coloneqq \dataoperator^{\pinv}\dataoperator\), do not commute in general, the two choices correspond to order in which these operations are computed.

The first option (A) is to re-use  \gls{POD} of the snapshot matrix \(\snapshotmatrix\) and choose \(\reduceddatadimension\) vectors from it, whether by choosing by their singular values or by performing an additional optimization as in~\Cref{sec:online-adapt}.
The span of chosen columns \(\mat{\Phi}\) is then additionally projected to the observation space by setting \(\datareduction_A = \projection_{\dataoperator}\mat{\Phi}_{
\reduceddatadimension}\), the \(\dataoperator\)-projection of the first \(\reduceddatadimension\) columns of the \(\snapshotmatrix\) based \gls{POD} modes. 

The second option (B) is to apply \gls{POD}
to the projected snapshot matrix
\(\projection_{\dataoperator} \snapshotmatrix\), compute the singular value decomposition \(\projection_{\dataoperator} \snapshotmatrix = \hat{\mat{\Phi}}\hat{\mat{\Sigma}}\hat{\mat{\Psi}}\), and then set \(\datareduction_B = \hat{\mat{\Phi}}_{\reduceddatadimension}\) by choosing \(\reduceddatadimension\) columns from \(\hat{\mat{\Phi}}\), the first \(\reduceddatadimension\) columns of the \gls{POD} modes of 
the \(\dataoperator\)-projection of the \(\snapshotmatrix\) snapshots. 

The difference between using \(\datareduction_A\) (the projected \gls{POD} vectors)  vs.~\(\datareduction_B\) (\gls{POD} vectors of projected snapshots) is subtle.
Comparing the projections onto spans of \(\datareduction_A\) and \(\datareduction_B\) shows that the (A) option is equivalent to the composition \(\projection_{\dataoperator} \projection_{\snapshotmatrix} \projection_{\dataoperator}\) while the projection in (B) option is equivalent to  \(\projection_{\dataoperator} \projection_{\snapshotmatrix}\), where in both cases \(\projection_{\snapshotmatrix}\) is the same projection on a \(\reduceddatadimension\)-dimensional subspace of the span of snapshots.
If there is a significant intersection between the cokernel of \(\dataoperator\) and the span of snapshots, the distinction between (A) and (B) is minor.
However, in the case that the column rank of \(\dataoperator\) is significantly smaller, or if a significant part of its range is orthogonal to the dominant subspace of snapshots, then the (A) case may overly restrict the size of the subspace in which assimilation is performed.
A more detailed discussion of this issue can be found in~\cite[\S 3--4]{MVV20}.
As the observation operators chosen below contain all variables, or a large subset of them, we do not expect that the distinction between \(\datareduction_A\) and \(\datareduction_B\) plays a major role.
For the remainder of this paper, we choose the (B) option for the computation of the data operator and drop the subscript \(\datareduction_A\) to simplify the notation.

Both dimensions \(\reducedmodeldimension\) and \(\reduceddatadimension\) can be chosen based on additional knowledge about the model or data space, e.g., if the model analysis suggests that dynamics collapses to a subspace of a known dimension.
Alternatively, the dimensions can be chosen in a data-driven way, by selecting the number of the retained POD vectors so that the  resulting approximation of matrices \(\snapshotmatrix\) and \(\dataoperator^{\pinv}\dataoperator\snapshotmatrix\) is within a certain distance (induced by \(\Vert \cdot \Vert_{F}\)) of the original matrices.
  We explore several versions of the data-driven approach in this work.

\subsection{Time-varying projection using a sliding window}
\glsreset{SWPOD}
The \gls{SWPOD} computes a time-dependent \gls{POD} projection operator based on a sliding subset of snapshots from the entire evolution.
The process of the \gls{SWPOD} starts by splitting the time interval \(\finaltime\)  into sub-intervals (\textit{windows}) \(W_i = [\timeindex_i,\timeindex_i + \windowsize]\), where \(\timeindex_i=[0,\, \windowshift,\, 2\windowshift, \dots]\).
Throughout this paper, we choose the window size to be twice the window shift, \(\windowsize = 2 \windowshift\), so that a generic time point \(\timeindex\) belongs to two consecutive windows \(W_{i}, W_{i+1}\).

The reduction matrices \(\statereduction_{W_{i}} = \begin{bmatrix}
  \leftsingularvector_{1} & \cdots & \leftsingularvector_{\reducedmodeldimension}
\end{bmatrix}\) are computed using \gls{POD} of windowed snapshot matrices \(\snapshotmatrix_{W_{i}}\) that are formed from columns of \(\snapshotmatrix\)  belonging to the window \(W_{i}\).
The choice of the dimension of the reduced space \(\reducedmodeldimension\), viz.~the rank of \(\statereduction_{W_{i}}\), is essential for the quality of the assimilation.
Below, we first present results where the \(\reducedmodeldimension\) is common to all windows.
Alternatively, the choice of the number of retained vectors can be adaptively chosen for each different window, resulting in time-varying choices \(\reducedmodeldimension(\timeindex)\) and \(\reduceddatadimension(\timeindex)\), as demonstrated in \cref{sec:adaptive-selection-of-parameters}.

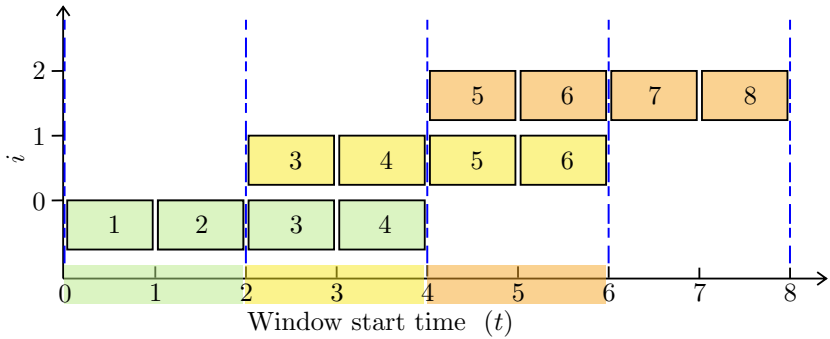
\begin{figure}
    \centering
\tikzset{every picture/.style={line width=0.70pt}} %
\begin{tikzpicture}[x=0.70pt,y=0.70pt,yscale=-0.70,xscale=0.70]
\draw  (60.86,250.42) -- (649.96,250.42)(60.86,40.5) -- (60.86,250.42) -- cycle (642.96,245.42) -- (649.96,250.42) -- (642.96,255.42) (55.86,47.5) -- (60.86,40.5) -- (65.86,47.5)  ;
\draw  [fill={rgb, 255:red, 184; green, 233; blue, 134 }  ,fill opacity=0.5 ] (64,190) -- (130,190) -- (130,228) -- (64,228) -- cycle ;
\draw  [fill={rgb, 255:red, 184; green, 233; blue, 134 }  ,fill opacity=0.5 ][line width=0.75]  (134,190) -- (200,190) -- (200,228) -- (134,228) -- cycle ;
\draw  [fill={rgb, 255:red, 184; green, 233; blue, 134 }  ,fill opacity=0.5 ] (204,190) -- (270,190) -- (270,228) -- (204,228) -- cycle ;
\draw  [fill={rgb, 255:red, 184; green, 233; blue, 134 }  ,fill opacity=0.5 ] (274,190) -- (340,190) -- (340,228) -- (274,228) -- cycle ;
\draw  [fill={rgb, 255:red, 248; green, 231; blue, 28 }  ,fill opacity=0.5 ] (204,140) -- (270,140) -- (270,178) -- (204,178) -- cycle ;
\draw  [fill={rgb, 255:red, 248; green, 231; blue, 28 }  ,fill opacity=0.5 ] (274,140) -- (340,140) -- (340,178) -- (274,178) -- cycle ;
\draw  [fill={rgb, 255:red, 248; green, 231; blue, 28 }  ,fill opacity=0.5 ] (344,140) -- (410,140) -- (410,178) -- (344,178) -- cycle ;
\draw  [fill={rgb, 255:red, 248; green, 231; blue, 28 }  ,fill opacity=0.5 ] (414,140) -- (480,140) -- (480,178) -- (414,178) -- cycle ;
\draw  [fill={rgb, 255:red, 245; green, 166; blue, 35 }  ,fill opacity=0.5 ] (344,90) -- (410,90) -- (410,128) -- (344,128) -- cycle ;
\draw  [fill={rgb, 255:red, 245; green, 166; blue, 35 }  ,fill opacity=0.5 ] (414,90) -- (480,90) -- (480,128) -- (414,128) -- cycle ;
\draw  [fill={rgb, 255:red, 245; green, 166; blue, 35 }  ,fill opacity=0.5 ] (484,90) -- (550,90) -- (550,128) -- (484,128) -- cycle ;
\draw  [fill={rgb, 255:red, 245; green, 166; blue, 35 }  ,fill opacity=0.5 ] (554,90) -- (620,90) -- (620,128) -- (554,128) -- cycle ;
\draw    (132,241) -- (132,249.8) ;
\draw    (202,241.04) -- (202,249.84) ;
\draw    (272,241) -- (272,249.8) ;
\draw    (342,240) -- (342,248.8) ;
\draw    (412,241) -- (412,249.8) ;
\draw    (482,241) -- (482,249.8) ;
\draw    (552,242) -- (552,250.8) ;
\draw    (622,242) -- (622,250.8) ;
\draw [color={rgb, 255:red, 0; green, 0; blue, 255 }  ,draw opacity=1 ][line width=0.75]  [dash pattern={on 3.75pt off 3pt on 7.5pt off 1.5pt}]  (202,50.5) -- (201.94,238.67) ;
\draw [color={rgb, 255:red, 0; green, 0; blue, 255 }  ,draw opacity=1 ][line width=0.75]  [dash pattern={on 3.75pt off 3pt on 7.5pt off 1.5pt}]  (342,50.5) -- (341.94,238.67) ;
\draw [color={rgb, 255:red, 0; green, 0; blue, 255 }  ,draw opacity=1 ][line width=0.75]  [dash pattern={on 3.75pt off 3pt on 7.5pt off 1.5pt}]  (482,50.5) -- (481.94,238.67) ;
\draw [color={rgb, 255:red, 0; green, 0; blue, 255 }  ,draw opacity=1 ][line width=0.75]  [dash pattern={on 3.75pt off 3pt on 7.5pt off 1.5pt}]  (622,50.5) -- (621.94,238.67) ;
\draw [color={rgb, 255:red, 0; green, 0; blue, 255 }  ,draw opacity=1 ][line width=0.75]  [dash pattern={on 3.75pt off 3pt on 7.5pt off 1.5pt}]  (62,50.5) -- (61.94,238.67) ;
\draw    (61,189.99) -- (52.02,190.09) ;
\draw    (61,140) -- (52.02,140.1) ;
\draw    (61,90) -- (52.02,90.1) ;
\draw  [color={rgb, 255:red, 0; green, 0; blue, 0 }  ,draw opacity=0 ][fill={rgb, 255:red, 184; green, 233; blue, 134 }  ,fill opacity=0.5 ] (62,240) -- (199.43,240) -- (199.43,270) -- (62,270) -- cycle ;
\draw  [color={rgb, 255:red, 0; green, 0; blue, 0 }  ,draw opacity=0 ][fill={rgb, 255:red, 248; green, 231; blue, 28 }  ,fill opacity=0.5 ] (202,240) -- (339.43,240) -- (339.43,270) -- (202,270) -- cycle ;
\draw  [color={rgb, 255:red, 0; green, 0; blue, 0 }  ,draw opacity=0 ][fill={rgb, 255:red, 245; green, 166; blue, 35 }  ,fill opacity=0.5 ] (342,240) -- (479.43,240) -- (479.43,270) -- (342,270) -- cycle ;

\draw (93,200) node [anchor=north west][inner sep=0.75pt]   [align=left] {$\displaystyle 1$};
\draw (160.5,200) node [anchor=north west][inner sep=0.75pt]   [align=left] {$\displaystyle 2$};
\draw (233.7,200) node [anchor=north west][inner sep=0.75pt]   [align=left] {$\displaystyle 3$};
\draw (302,200) node [anchor=north west][inner sep=0.75pt]   [align=left] {$\displaystyle 4$};
\draw (233.1,150) node [anchor=north west][inner sep=0.75pt]   [align=left] {$\displaystyle 3$};
\draw (303,150) node [anchor=north west][inner sep=0.75pt]   [align=left] {$\displaystyle 4$};
\draw (373,150) node [anchor=north west][inner sep=0.75pt]   [align=left] {$\displaystyle 5$};
\draw (510.2,100) node [anchor=north west][inner sep=0.75pt]   [align=left] {$\displaystyle 7$};
\draw (442.9,150) node [anchor=north west][inner sep=0.75pt]   [align=left] {$\displaystyle 6$};
\draw (584.1,100) node [anchor=north west][inner sep=0.75pt]   [align=left] {$\displaystyle 8$};
\draw (442.8,100) node [anchor=north west][inner sep=0.75pt]   [align=left] {$\displaystyle 6$};
\draw (373.7,100) node [anchor=north west][inner sep=0.75pt]   [align=left] {$\displaystyle 5$};
\draw (125,252) node [anchor=north west][inner sep=0.75pt]   [align=left] {$\displaystyle 1$};
\draw (195,252) node [anchor=north west][inner sep=0.75pt]   [align=left] {$\displaystyle 2$};
\draw (265,252) node [anchor=north west][inner sep=0.75pt]   [align=left] {$\displaystyle 3$};
\draw (335,252) node [anchor=north west][inner sep=0.75pt]   [align=left] {$\displaystyle 4$};
\draw (405,252) node [anchor=north west][inner sep=0.75pt]   [align=left] {$\displaystyle 5$};
\draw (475,252) node [anchor=north west][inner sep=0.75pt]   [align=left] {$\displaystyle 6$};
\draw (55,253) node [anchor=north west][inner sep=0.75pt]   [align=left] {$\displaystyle 0$};
\draw (545,252) node [anchor=north west][inner sep=0.75pt]   [align=left] {$\displaystyle 7$};
\draw (615,252) node [anchor=north west][inner sep=0.75pt]   [align=left] {$\displaystyle 8$};
\draw (345.2,286.17) node   [align=left] {\begin{minipage}[lt]{12.92pt}\setlength\topsep{0pt}
$ $
\end{minipage}};
\draw (200.66,272.84) node [anchor=north west][inner sep=0.75pt][align=left]{Window start time $\ (\timeindex)$};
\draw (35,181.7) node [anchor=north west][inner sep=0.75pt]   [align=left] {$\displaystyle 0$};
\draw (35,131.7) node [anchor=north west][inner sep=0.75pt]   [align=left] {$\displaystyle 1$};
\draw (35,80.7) node [anchor=north west][inner sep=0.75pt]   [align=left] {$\displaystyle 2$};
\draw (15,163.95) node [anchor=north west][inner sep=0.75pt]  [rotate=-269.07] [align=left] {$i$};
\end{tikzpicture}
\caption{Illustration of the window overlaps in \gls{SWPOD} for a data set containing 8 snapshots, with three windows of size \(\windowsize=4\) and window shift \(\windowshift=\windowsize/2=2\).}
\label{fig:windows}
\end{figure}

\Cref{fig:windows}, illustrates the \gls{SWPOD} with windows containing only four snapshots \(\snapshotmatrix=
\begin{bmatrix}
\state_{1}&\state_{2}&\state_{3}&\state_{4}
\end{bmatrix}\).
Three windows are shown in the graph with window sizes \(\windowsize=4\) and window shift \(\windowshift=\windowsize/2\).
Two new snapshots are added at time \(\windowshift+1\), and the oldest snapshots are dropped.
\gls{POD} is computed for windowed \(\snapshotmatrix_{[\timeindex, \timeindex+\windowsize]}\); it is updated every \(\windowshift\) steps and singular values \(\sigma_{m}\) and corresponding modes \(\leftsingularvector_{m}\) are computed.

During the online phase of the assimilation, at every time instance \(\timeindex\), the algorithm chooses between the reduction matrices \(\statereduction_{W_i}\) or \(\statereduction_{W_{i+1}}\) for two consecutive windows which both contain \(\timeindex\).
In all computations presented here, we choose the later window, with index \(i(\timeindex) = \floor{\timeindex/\windowshift}\).

\subsection{Strategies for choosing parameters of the reduction}
\label{sec:adaptive-selection-of-parameters}

We introduced the \gls{SWPOD} reductions to react to regime changes in the model dynamics that may happen during the assimilation period.
The first strategy we present here is the offline fixed dimension where the parameters of model reduction \(\reducedmodeldimension\) and data reduction \(\reduceddatadimension\) are set to a fixed value per window assimilation.

\begin{description}
\item[Offline Fixed Dimension.]
The offline fixed dimension strategy selects a fixed size of the reduced physical model dimension \(\reducedmodeldimension\) and data model dimension \(\reduceddatadimension\) based on the offline-computed digital twin simulation, i.e., before the assimilation/analysis loop begins.
The offline techniques have mode selection at fixed times which we assume here to be equally spaced.
This treats mode selection as a relatively scarce (or expensive) resource that cannot be used on demand but only at scheduled times.
This can clearly cause difficulties if there are changes in the dynamics but the modes are not updated.
The techniques developed here also apply when mode selection times are found adaptively.
\end{description}
In case of a significant change in the dimension of the subspace in which dynamics are concentrated, the dimensions \(\reducedmodeldimension\) and \(\reduceddatadimension\) can be adapted in time as well.
The adaptive mode selection strategy aims to improve the assimilation and avoid overfitting and underfitting that might happen with fixed, reduced dimensions.
Using the sliding window POD, the number of retained vectors can be tailored for every window, resulting in time-varying choices for the number of retained vectors.

We present here the strategies for determining \(\reducedmodeldimension(\timeindex)\) and \(\reduceddatadimension(\timeindex)\) adaptively.
\begin{description}
  \item[Offline Adaptive Dimension.]
The offline strategy is computes both \(\reducedmodeldimension(\timeindex)\) and \(\reduceddatadimension(\timeindex)\) based on the offline-computed digital twin simulation, i.e., before the assimilation/analysis loop begins.
The number of modes is determined by retaining a sufficient fraction of \gls{POD} modes needed to compress the block of snapshots belonging to the window to a pre-determined tolerance.
This produces a piecewise-constant change in both reduction dimensions.
The reduction matrices \(\statereduction_{\timeindex}\) and \(\datareduction_{\timeindex}\) are constructed by choosing the most dominant \(\reducedmodeldimension(\timeindex)\) and \(\reduceddatadimension(\timeindex)\) \gls{POD} vectors corresponding to the selected window \(i(t)\).
  \item[Online Sparse Subspace.]
  Model reduction dimension \(\reducedmodeldimension(\timeindex)\) is computed as above.
  In contrast to the above strategies, the data reduction operator is not constructed from the most dominant \gls{POD} vectors.
  Instead, this strategy employs sparse selection (\(\ell^1\) minimization) to choose a subset of \gls{POD} vectors to compress the incoming measurement within a specified tolerance.
\end{description}

\subsubsection{Offline adaptivity}\label{sec:offline-adapt}
Recall, to achieve a model reduction of the \(\snapshotmatrix_{[\timeindex, \timeindex+\windowsize]}\), we form the reduction matrix \( \mat{\statereduction}_{[\timeindex, \timeindex+\windowsize]} =
  \begin{bmatrix}
    \leftsingularvector_{1} & \cdots & \leftsingularvector_{\reducedmodeldimension}
  \end{bmatrix},\) \eqref{eq:POD-projection} where we choose
  \(\reducedmodeldimension \leq \modeldimension\) (and hopefully \(\reducedmodeldimension \ll \modeldimension\)), and then perform the projection on the set of the POD basis \(\begin{bmatrix}
    \leftsingularvector_{1} & \cdots & \leftsingularvector_{\reducedmodeldimension}
  \end{bmatrix}\).
  The choice of the rank of the projection (number of retained POD basis elements) can be crucial.

With the benefit of the sliding window POD, the choice of the number of retained vectors can be performed for each different window, resulting in time-varying choices \(\reducedmodeldimension(\timeindex)\) and \(\reduceddatadimension(\timeindex)\).
Since performing such a prior choice is unfeasible, we choose the dimensions automatically based on a tolerance parameter \(\tol\).

  Assuming that the singular values are ordered in decreasing order, \(\sigma_{i}\geq \sigma_{j}\) for \(i<j\), the order
\begin{equation}
\label{eq:tol}
    r=\min \{ k :\sum_{m=1}^{k}\sigma_m^2\leq \tol \sum_{m=1}^{M}\sigma_m^2\}
\end{equation}
is the smallest number of modes required to retain the fraction \(\tol \in [0,1]\) of the total \(\norm{\cdot}_F\)-norm of the matrix.

We employ two tolerance parameters,  \(\tol_{\reducedmodeldimension}\) and \(\tol_{\reduceddatadimension}\) in place of \(\tol\) in \eqref{eq:tol}, controlling the degree of reduction in the model and data spaces respectively, and apply the selection procedure separately to \gls{POD} decompositions of windowed snapshot matrix \(\snapshotmatrix_{[\timeindex, \timeindex+\windowsize]}\) and its restriction \(\dataoperator^{\pinv}\dataoperator\snapshotmatrix_{[\timeindex, \timeindex+\windowsize]}\) to the cokernel of the observation operator \(\dataoperator\).

As a result of this process, \(\statereduction_{\timeindex}\), \(\reducedmodeldimension(\timeindex)\), \(\datareduction_{\timeindex}\), and \(\reduceddatadimension(\timeindex)\) are piecewise-constant functions, changing only when the choice of the time window is changed, and can be fully determined before the analysis loop begins.

\subsubsection{Online Adaptivity}\label{sec:online-adapt}
The online adaptive method aims to determine \(\reduceddatadimension(\timeindex) \leq \reducedmodeldimension(\timeindex)\) for each time step based on the just observed measurement.
The \(\statereduction_{\timeindex}\) and  \(\reducedmodeldimension(\timeindex)\) are determined as in \cref{sec:offline-adapt}; then \(\datareduction_{\timeindex}\) is chosen as the subset of columns of \(\statereduction_{\timeindex}\) that best approximate \(\dataoperator^{\pinv}\data_{\timeindex}\).
The dimension \(\reduceddatadimension\) and the \(\datareduction\) operators are used only in the weight-update step of the \gls{OP-PF}, which we can compute the \(\datareduction_\timeindex\) and choose the size  \(\reduceddatadimension(\timeindex)\)  online for each observation.
Reducing \(\reduceddatadimension\) significantly is associated with staving off the particle degeneracy, which decreases the effectiveness of the particle filter assimilation strategy.

The proposed rule chooses the columns \(\datareduction\) as the subset of columns of \(\statereduction\) by performing a sparse \(\ell^2\) regression (\emph{lasso} \cite{tibshirani1994}) of the incoming measurements onto the POD basis.
The regression coefficients \(\vec{c}^\ast\) are computed by the minimization
\begin{equation}%
\vec{c}^\ast= \argmin_{\vec{c} \in \mathbb{R}^{\reducedmodeldimension}}%
\left[%
\norm{\statereduction_{\timeindex} \mat{c}-\dataoperator^{\pinv}\data_\timeindex}_{2}+\tol_{D}\norm{\mat{c}}_{1}%
\right]%
\label{eq:lasso-regression}%
\end{equation}
where the term \(\norm{\vec{c}}_1\) is responsible for promoting sparsity of the coefficient vector \(\vec{c}\).
The indexes of non-zero coefficients \(\vec{c}^\ast\) then correspond to those columns in \(\statereduction\) that are used to construct \(\datareduction\).
The target of regression \(\dataoperator^{\pinv}\data_\timeindex\)
promotes retaining only POD modes that are in the cokernel of \(\dataoperator\), i.e., those POD modes whose magnitude is observable under the action of the \(\dataoperator\).
The \emph{lasso} optimization is the convex relaxation of selecting \emph{best subset} of columns of \(\statereduction_{\timeindex}\) to approximate \(\dataoperator^{\pinv}\data_{\timeindex}\).
More on this can be found in standard references on statistical inference, for example~\cite{hastie2009elements, tibshirani1994}.

The parameter \(\tol_D \in [0,1)\) is tunable.
Setting it to \(\tol_D = 0\) results in a ``standard'', i.e., non-sparse, \(\ell^2\) regression, which generically sets \(\datareduction_{\timeindex} = \statereduction_{\timeindex}\) (and consequently \(\reduceddatadimension=\reducedmodeldimension\)).
Increasing \(\tol_D\) results in the increased importance of the sparsity-promoting term \(\norm{\vec{c}}_1\), which reduces \(\reduceddatadimension\) below \(\reducedmodeldimension\).
Computationally, most numerical packages have algorithms that efficiently solve this regression and the impact on the runtime of the analysis procedure is minimal.

Both non-regularized (\(\tol_D=0\)) and \(\ell^1\)-regularized (\(\tol_D > 0\)) regressions can be given a bayesian interpretation.
Bayesian regression equivalent to the given procedure  consists in assuming a prior distribution for each coefficient being tuned, computing a posterior distribution based on the data, that is the term \(\dataoperator^{\pinv}\data_\timeindex\) in \eqref{eq:lasso-regression}, and then computing the mode of the posterior as the value for each parameter.
The non-regularized case corresponds to taking the uniform distribution as the prior for each parameter; the regularized case instead uses the bi-exponential (Laplace) distribution \(\sim \exp(-\tol_D \abs{c_i})\) as the prior~\cite{Park2008}.
Replacing the \(\ell^1\) norm with the \(\ell^2\) norm in \eqref{eq:lasso-regression} would correspond to the normal distribution being used as a prior.
Since the Laplace distribution is sharply peaked at zero, it is more likely to yield zero-valued coefficients than either normal or uniform distributions, resulting in overall sparser coefficient vectors \(\vec{c}\).

This procedure results in a piecewise-constant \(\statereduction_{\timeindex}\) that is determined offline (before the analysis loop) and changes only when the window changes.
The \(\datareduction_{\timeindex}\) varies in each observation step, as it is computed depending on the incoming data \(\data_{\timeindex}\).

\glsreset{L96}
\section{Numerical evaluation of the assimilation}\label{sec:results}

We evaluate the described assimilation procedure on the often-used \gls{L96}~\cite{Lorenz96, ott2004local}.
It is a system of autonomous, nonlinear, ordinary differential equations parametrized by the dimension of the system and the value of the forcing parameter, first developed as a simplified model of global horizontal circulation of the atmosphere.
Depending on the combination of these values, the system can exhibit behaviors ranging from steady states and regular traveling waves to fully-developed chaos.

In the first phase of evaluation, we demonstrate the effect that the recomputation of basis has on the assimilation in both regular and chaotic regimes, even when the reduction dimension is not tailored.
In the second phase, we evaluate the adaptive strategies from~\cref{sec:adaptive-selection-of-parameters} that can be used to change how many basis elements and which basis elements are used in the reduction scheme.

Overall, we demonstrate that the sliding window with the adaptive tuning of the reduction operator brings a significant improvement in efficacy as compared to the time-invariant reduction operator.

\subsection{Lorenz '96 model}\label{sec:l96-model}

The state of the model is a vector \(\state = (u_{i})_{i=1}^\modeldimension\) of an arbitrary dimension \(\modeldimension\),
evolving according to the ordinary differential equation 
\begin{equation} \label{eq:l96}
    \dot u_{i}  = u_{i-1}\left(u_{i+1}-u_{i-2}\right) -u_{i}+ F, \quad i=1,\dots,\modeldimension,
\end{equation}
with the periodic boundary condition \(u_{i} \equiv u_{i+\modeldimension}\).
The parameter \(F\) determines if the evolution will be qualitatively regular or chaotic.

The discrete-time evolution map \(\modelupdate\) used to formulate the assimilation scheme in~\eqref{eq:physical-model} is computed by solving~\eqref{eq:l96} using an adaptive Runge--Kutta scheme (Dormand--Prince pair used by MATLAB's \texttt{ode45}) with the solution resampled at multiples of the observation time \(\timestepobs\).

\begin{figure}[htb]
  \centering
  \includegraphics[width=0.70\textwidth]{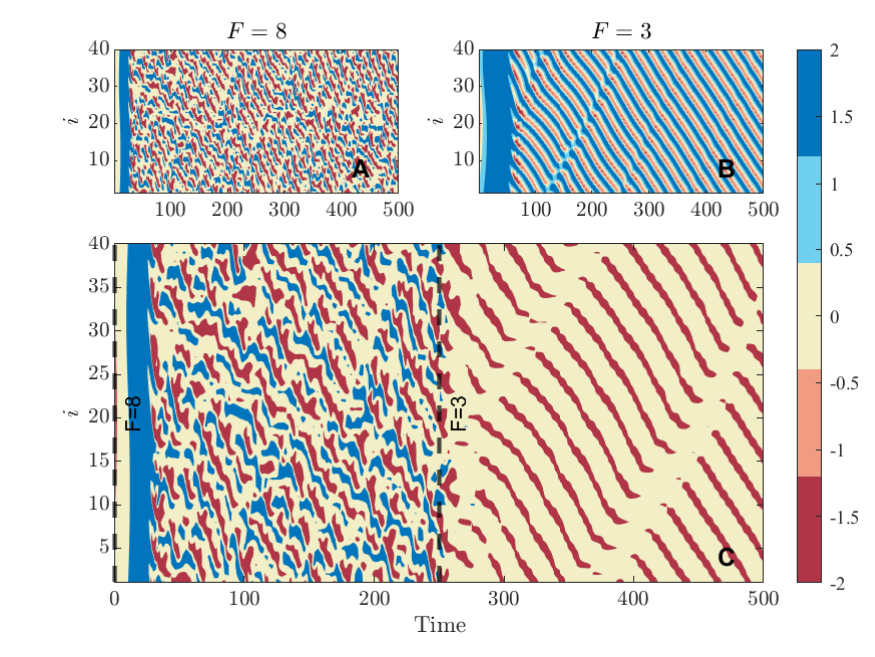}
  \caption{Solutions \(\state(\timeindex)\) of the \gls{L96}.
   Figures (A) and (B) show the chaotic and regular structure for \(F=8\) and \(F=3\) respectively.
Figure (C) shows the solution in which regime change \(F=8\rightarrow F=3\) (chaotic to regular) occurs at \(\timeswitch=\finaltime/2\).
}
\label{fig:l96-space-time}
\end{figure}

To produce a regime change in the model, the forcing parameter \(F\) is replaced by a time-varying function.
In all our simulations,  \(F\) changes discontinuously between \(F=3\) and \(F=8\) at one, or more, time instances \(\timeindex = \timeswitch\).
\Cref{fig:l96-space-time} illustrates the space-time behavior of a typical solutions of \gls{L96} in a described configuration.
In all cases, the initial condition is \(u_{m}(0) = \cos(2 \pi m/M),\ m=1,\dots, M\).

The model-reduction projections \(\projection\) are in this work generated using singular vectors of \gls{POD} for a single simulation of model equations.
To demonstrate the degree to which the range of the projection \(\projection\) conforms to alternative realizations of the model evolution, we perform the following calculations.

First, we generate \(\statetruth_\timeindex\) by integrating the (deterministic) model equations~\eqref{eq:l96} for \(\modeldimension = 400\), where we change the value of \(F = 8 \to 3\) during the simulation, therefore triggering the regime change.
This evolution is then used to generate a projection matrix \(\projection\) of order \(\reducedmodeldimension=390\) as explained in~\Cref{sec:dimension-reduction}.
Since \(\projection\) is generated by retaining \(\reducedmodeldimension\) dominant \gls{POD} vectors, we can use the singular values to estimate the time-integrated error \( \norm{(\id - \projection)\statetruth}_p / \norm{\statetruth}_p\), for induced norm \(p = 2\) or Frobenius norm \(p=F\).
However, we are here additionally interested in how the snapshot error \( E^{\text{truth}}_\timeindex \coloneqq \norm{(\id - \projection)\statetruth_\timeindex}_2 / \norm{\statetruth_\timeindex}_2\) evolves in time.

Second, starting from the same initial condition \(\state_0 =\statetruth_0\), we integrate the model equations with the stochastic error~\eqref{eq:physical-model} to produce the evolution \(\state_\timeindex\).
We then compute the relative snapshot error \( E_\timeindex \coloneqq \norm{(\id - \projection)\state_\timeindex}_2 / \norm{\state_\timeindex}_2\) using the same projection matrix \(\projection\) computed from the ``noiseless'' evolution \(\statetruth_\timeindex\).

Finally, we apply the \gls{ProjOPPF}~\cite{albarakati2021model} to assimilate observations generated using \(\statetruth_\timeindex\) to produce the ensemble estimate \(\stateens_\timeindex\) as the weighted ensemble mean~\eqref{eq:ensemble-estimate}.
The same projection \(\projection\) is used in the algorithm, with the ensemble size \(L =20\).
Again, we compute the relative snapshot error \( E^{\text{ens}}_\timeindex \coloneqq \norm{(\id - \projection)\stateens_\timeindex}_2 / \norm{\stateens_\timeindex}_2\).

\begin{figure}[htb]
\centering
\begin{subfigure}{0.49\textwidth}
  \centering
\includegraphics[width=.90\linewidth]{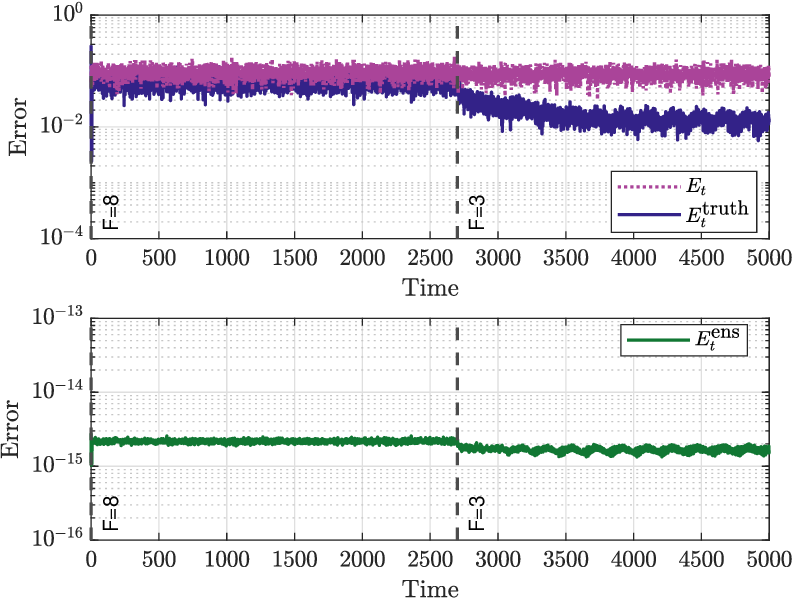}
  \caption{Relative snapshot error}
  \label{fig:E}
\end{subfigure}%
\begin{subfigure}{0.49\textwidth}
  \centering
\includegraphics[width=.90\linewidth]{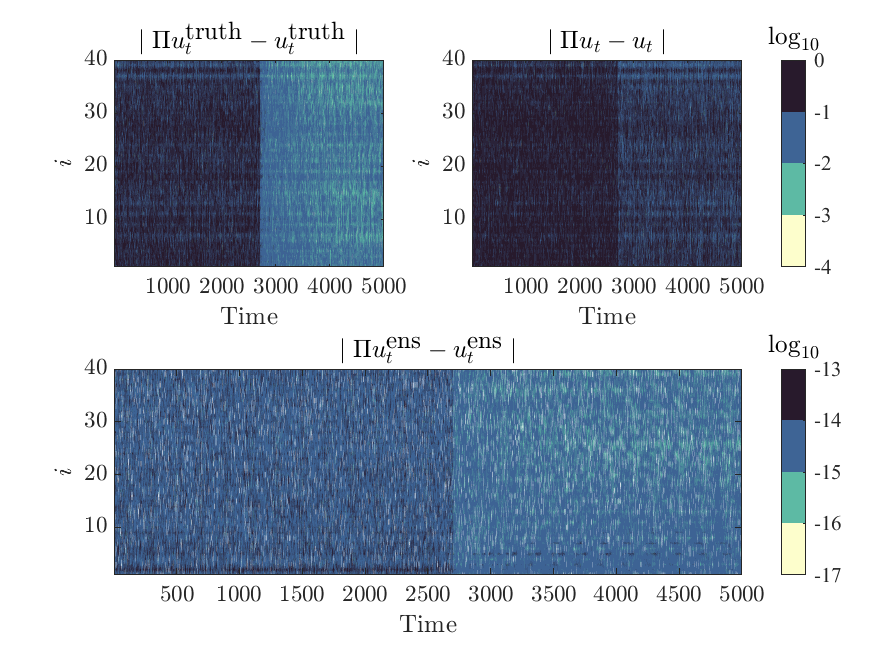}
  \caption{Pointwise error difference }
  \label{fig:logE}
\end{subfigure}
\caption{Compatibility between the subspace of training data and the ensemble estimate. Figure (a) shows the time traces for the three errors  \( E^{\text{truth}}_\timeindex \coloneqq \norm{(\id - \projection)\statetruth_\timeindex}_2 / \norm{\statetruth_\timeindex}_2\), \( E_\timeindex \coloneqq \norm{(\id - \projection)\state_\timeindex}_2 / \norm{\state_\timeindex}_2\), and \( E^{\text{ens}}_\timeindex \coloneqq \norm{(\id - \projection)\stateens_\timeindex}_2 / \norm{\stateens_\timeindex}_2\). Figure (b) shows the pointwise difference between the projected and unprojected solutions. See \cref{sec:l96-model} for more details.}
\label{fig:E-truth}
\end{figure}

\Cref{fig:E-truth} shows the time traces for each of the three errors \(E^{\text{truth}}_\timeindex\), \(E_\timeindex\), and \(E^{\text{ens}}_\timeindex\), as well as the pointwise difference between the projected and unprojected solutions.
The relative and pointwise errors for \(\state^{\text{truth}}\) are small, but not negligible, representing the truncation error of the singular value to retain the dominant \(\reducedmodeldimension< \modeldimension\) basis vectors. 
The noisy evolution \(\state_\timeindex\) has slightly larger errors as neither the noise, or nonlinear evolution function, are constrained to be in the range of \(\projection\).
Since the ensemble estimate \(\stateens_\timeindex\) is computed in the reduced space, and reconstructed using the orthonormal system spanning the range of \(\projection\), the \(\projection \stateens_\timeindex\) is constrained to be in the range of \(\projection\). 
As a result, the relative and pointwise difference errors for \(\stateens_\timeindex\) are on the order of machine precision (e.g., \(\sim 10^{-15}\) ).

\Cref{fig:E-truth} demonstrates that when a time-invariant projection \(\projection\) is derived from a solution with variable spatial complexity, as \(\state^{\text{truth}}_\timeindex\), even that same solution can be significantly (mis)aligned with the range of \(\projection\).

This leads us to consider time-\emph{varying} projections, computed by applying \gls{POD} to segments of the trajectory (sliding data windows), expecting that regime change points would be restricted to only a small proportion of such windows, leading to a more stable and predictable fit between the solution and the used projection.
To demonstrate that the degree of complexity of the solutions changes within the two regimes, we compute the \gls{POD} decomposition within each regime.
Generally speaking, order reduction techniques are effective when a relatively small number of modes are capable of reproducing the dynamics.
For \gls{POD}, this is indicated by the magnitude of singular values \(\sigma_{m}\), ordered in descending order.
The singular vectors (modes) associated with small \(\sigma_{m}\) are thought of as less important for reconstructing the solution, with accuracy measured by the \(\ell_{2}\)-norm.
Therefore, regimes in which \(\sigma_{m}\) quickly drop off are interpreted as simpler, and we expect order reduction techniques based on \gls{POD} to perform better.

\begin{figure}[ht]
\begin{subfigure}[t]{0.32\linewidth}\centering
\includegraphics[width=\textwidth]{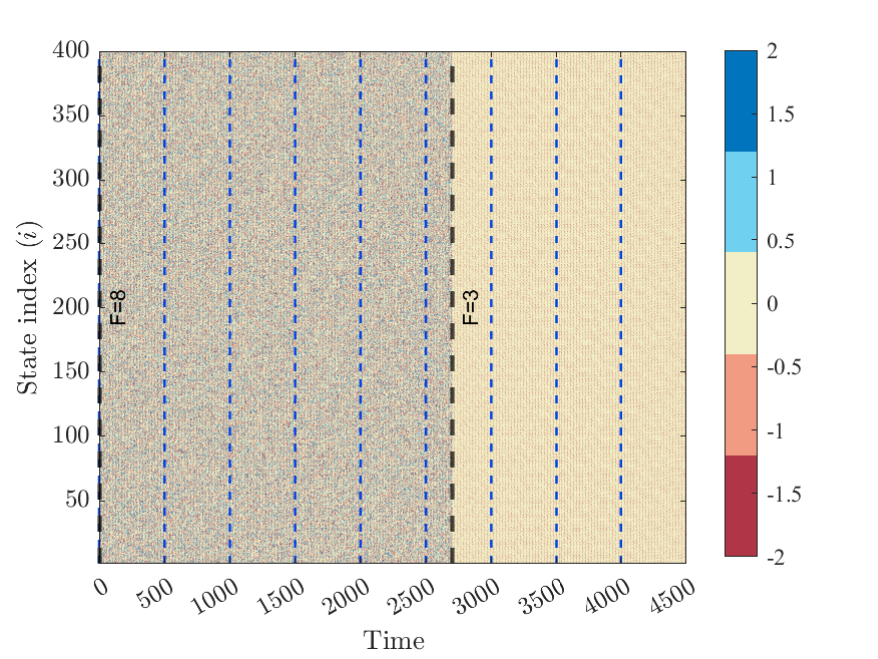}
\caption{Spatiotemporal evolution with the switching of windows indicated by vertical lines.}
\label{fig:L96-swpod-evolution}
\end{subfigure}
\begin{subfigure}[t]{0.32\linewidth}\centering
\includegraphics[width=\textwidth]{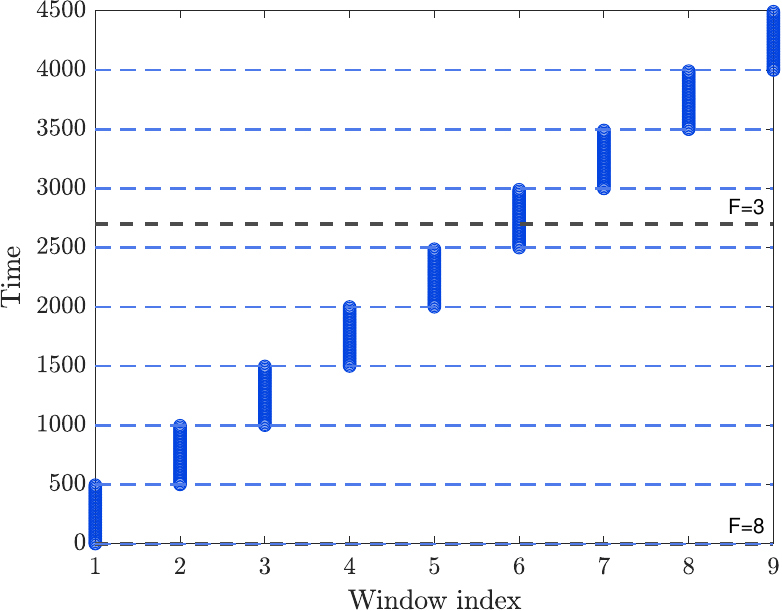}
  \caption{Window choice.}
  \label{fig:L96-swpod-window-choice}
  \end{subfigure}
\begin{subfigure}[t]{0.32\linewidth}\centering
\includegraphics[width=\textwidth]{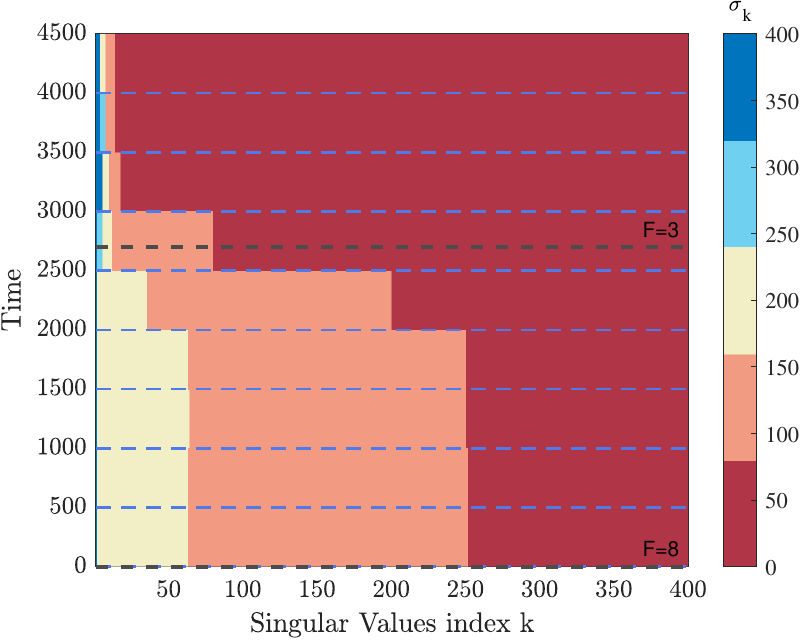}
\caption{Magnitude of singular values of \gls{SWPOD} for the time windows chosen at each time step.}
\label{fig:L96-swpod-singular-values}
\end{subfigure}
\caption{Spatiotemporal evolution and singular values of the \gls{SWPOD} for the \gls{L96} of dimension \(\modeldimension=400\) with the changing forcing parameter \(F=8\rightarrow F=3\) at \(\timeswitch=2700\).
\gls{SWPOD} is computed for \(9\) windows of size \(\windowsize=1000\), shifted by \(\windowshift=500\).
Panel~\ref{fig:L96-swpod-singular-values} shows that in the chaotic regime (\(F=8\)) there are far more non-negligible singular values than in the regular regime (\(F=3\)).
}

\label{fig:SWPOD_setup}
\end{figure}

\Cref{fig:SWPOD_setup} shows the setup of the \gls{SWPOD} for \gls{L96} model with model and data dimensions \(\modeldimension=\datadimension
=400\) with the changing forcing parameter \(F=8\rightarrow F=3\) at \(\timeswitch=\finaltime/2=2500\) indicated by vertical dashline.
The time axis is distributed between time windows as in the rest of the manuscript: the subsequent windows are always overlapped by half 
(\(\windowshift=\windowsize/2\)).
\Cref{fig:L96-swpod-singular-values} shows that the singular values computed for the \gls{SWPOD} drop off much more slowly in the first half of the evolution (regular regime), compared to the second half (chaotic regime).
As a result, we expect that during the regular regime the order reduction techniques will perform well even when \(\reducedmodeldimension \ll \modeldimension\).

\glsreset{ESS}
\subsection{Experimental setup}\label{sec:exp-setup}

The experiments in which we evaluate the data assimilation share the following setup. 
We use \gls{L96} with the state space of dimension \(\modeldimension=400\).
The particles are initialized with equal weights \((1/\totalparticles)\) with a fixed number of particles, \(\totalparticles=20\).
 
The projected \gls{ESS} is calculated \eqref{eq:ESS1} and projected resampling, using the scheme described in~\Cref{sec:ProjOPPF} is performed  with the spread of particles governed by \cref{resampling}, where the proportion of resampling variance inside the projection subspace is always taken to be \(\alpha = 0.99\) using the multinomial resampling scheme (see, e.g., \cite{Douc}).
Total resampling variances \(\omega = 10^{-6}\) when \(\ess<\frac{1}{2}\totalparticles\).
The physical and data models error covariances are fixed to be 
\(\modelerrorcovariance=1\,\id\) and \(\dataerrorcovariance = 0.01\,\id\), respectively.
The standard deviation of observation error of $0.1$ is included for comparison in the figures when reporting on \(\rmse\).
The observations are computed at every step, yielding the effective time step of assimilation \(\timestepobs = 1\).
The assimilation is performed over \(5000\) observation times.

The  ``truth'', \(\statetruth_{\timeindex}\) is determined by running a single simulation of the model.
The noisy observations of the \(\statetruth\) evolution computed using the data model \eqref{eq:data} are fed into the assimilation process.

\glsreset{RMSE}\glsreset{RESAMP}\glsreset{SWPOD}
We refer to the calculation of \gls{POD} without employing the sliding-window as NON-SW-POD and we use it as the baseline case for comparison with various described variants of \gls{SWPOD}.
For each assimilation algorithm, the success of assimilation can be measured by how closely the ensemble estimate \(\stateens\) matches \(\statetruth\).
Numerically, we quantify this match by several measures.
\begin{itemize}
\item Pointwise difference \(\Delta \in \mathbb{R}^\modeldimension\) is used to show the spatial distribution  of the error 
\begin{equation}
  \Delta \coloneqq \abs{\statetruth - \stateens},
  \label{eq:Error}
\end{equation}
\item \gls{RMSE}, a scalar measure of the difference between the estimate of a state and the true state,
  \begin{equation}
  \rmse \coloneqq \norm*{\Delta}_2/\sqrt{\modeldimension},\label{eq:rmse}
\end{equation}
where \(\modeldimension\) denotes the model dimension, expresses the quality of the estimate using a scalar.
\item \glsreset{ESS}\gls{ESS} 
measures the spread of the weights across particles, and is an indicator of the particle filter collapse as described in~\Cref{sec:ProjOPPF}
\begin{equation}
\ess = \left[{\sum_{\particleindex=1}^\totalparticles \left(\weight^{\particleindex}\right)^2}\right]^{-1}
\leq {\tt threshold}.
\label{eq:ESS}\end{equation}
\item The \gls{RESAMP}  measures the number of times the particle population had to be resampled.
We report the moving mean of twenty consecutive \gls{RESAMP} values.
\end{itemize}

Better performance is implied by higher values of \gls{ESS} and lower values of all other indicators.
We want the \gls{ESS} to stay above the threshold \(\frac{1}{2}\totalparticles\) to avoid frequent resampling.
In some experiments, we calculate the moving minimum over time of \gls{ESS} where each minimum is calculated over a moving window of length \(20\).
We compare some of the results to the unprojected particle filter (\gls{I-Proj}), or equivalently the use of identity matrices \(\statereduction=\mat{I}\), \(\datareduction=\mat{I}\).

The experiments (1 -- 4) evaluate two types of adaptive techniques described in \cref{sec:offline-adapt} and \cref{sec:online-adapt}.
The experiments' parametrization details are given in \cref{tab:online-exp}.

\begin{table}[htb]
  \centering
  \rowcolors{1}{}{white!90!black}\small
  \begin{tabular}{>{\bfseries\normalsize}c | >{\centering\arraybackslash}p{0.5in}  | c | >{\centering\arraybackslash}p{0.75in} | c  | c | c | c}
    {Exp.}  &  \(F\) &\(\finaltime\)& \(\timeswitch\) & \(\timestepobs\) &  \(\dataoperator\) & Adaptivity \\ \hline
    1 & \(8\rightarrow 3\) &\(5000\)& \(2700\) & \(1\) &  \(\id\) & offline \\
    2 & \(8\rightarrow 3\) &\(5000\)& \(2700\) & \(1\) &  \(\id\) & online \\
    3 &  \(8\rightarrow 3\) &\(1000\)& \(500\) &  0.025 -- 1 & \(\downsamp{2}\) & online \\
    4 & \(8\rightarrow 3 \to\) \(\dots \to  8\) &  \(10000\) & \(\approx\) every 2500 & \(1\) &  \(\id\) & online \\
    \end{tabular}
    \caption{Parametrization of offline and online adaptivity experiments where \(F\) is time-varying forcing for \gls{L96} that changes the behavior of the dynamics from regular to chaotic interchangeably at \(\timeswitch\), and \(\windowshift\) is the window time shift.
The column \(\dataoperator\) indicates the observation operator used: \(\id\) when all variables are observed, while \(\downsamp{2}\)  is canonical projections onto every-other state variable.
For all experiments, window shift is \(\windowshift = 500\), the model error covariance is \(\modelerrorcovariance=1\) and the data error covariance is \(\dataerrorcovariance=0.01\).}\label{tab:online-exp}
  \end{table}
\subsection{Summary of experimental results}
\label{sec:EXP-results}

\begin{figure}[ht!]
 \centering
   \begin{subfigure}[t]{0.32\linewidth}\centering
   \includegraphics[width=\textwidth]{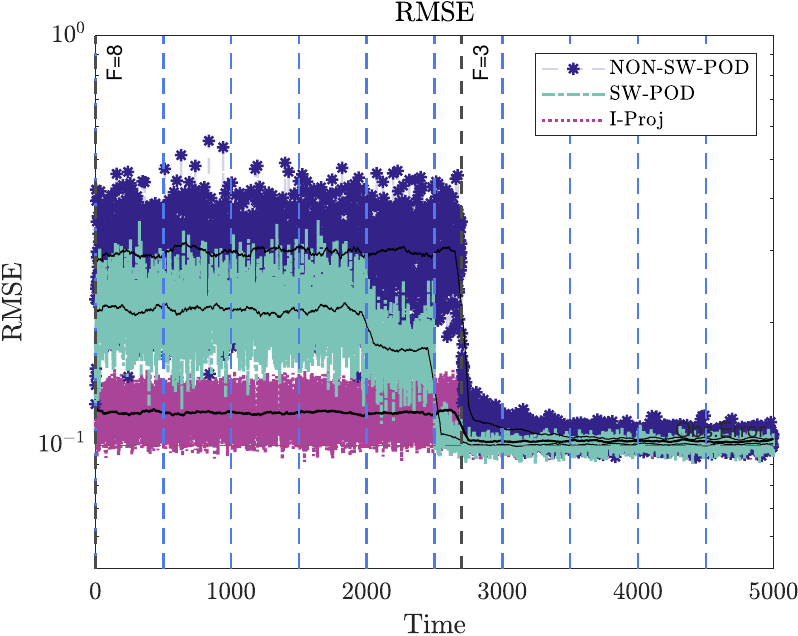}
   \caption{\gls{RMSE}, \(\reducedmodeldimension =390\)}
   \end{subfigure}
   \begin{subfigure}[t]{0.32\linewidth}\centering \includegraphics[width=\textwidth]{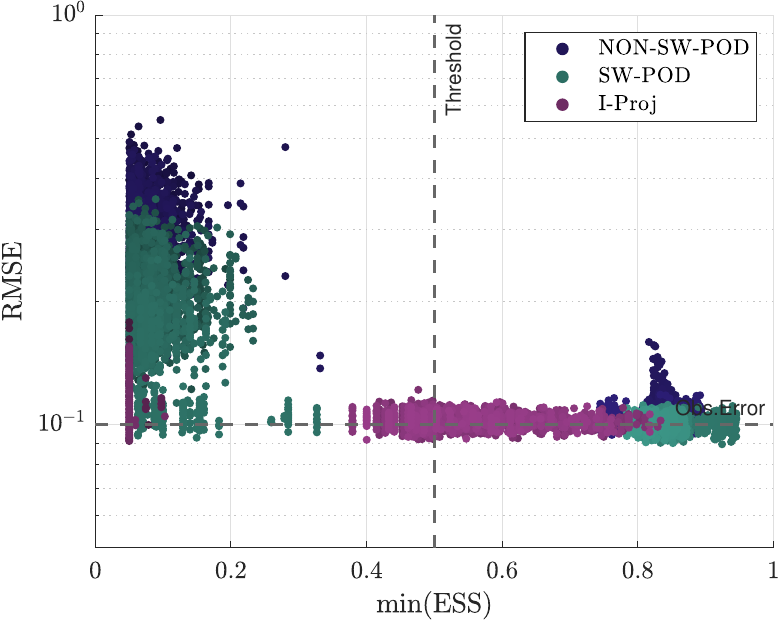}
   \caption{Trade-off between \gls{RMSE} and min(\gls{ESS}), \(\reducedmodeldimension =390\)}
   \end{subfigure}
   \begin{subfigure}[t]{0.32\linewidth}\centering
   \includegraphics[width=\textwidth]{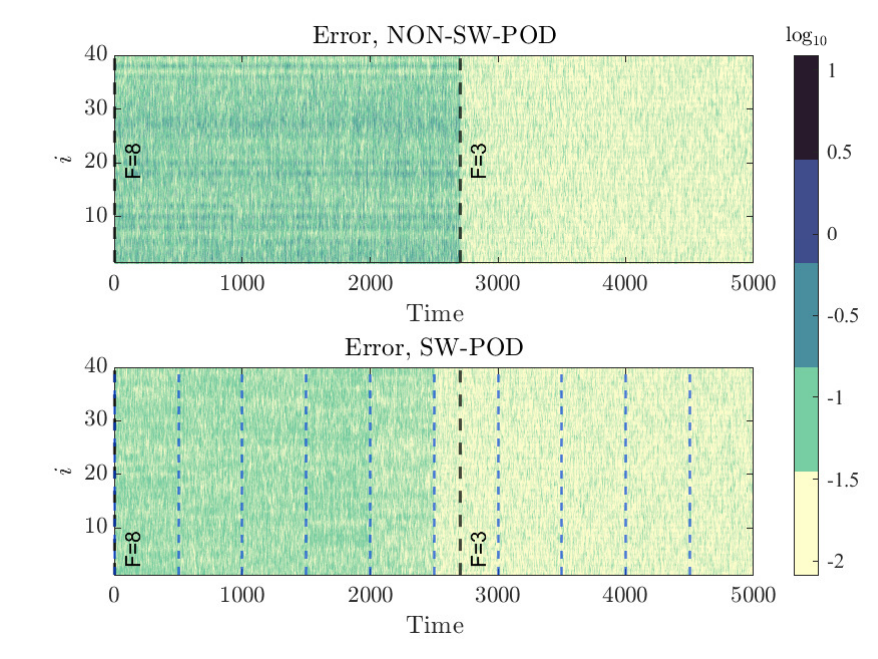}
   \caption{Error differences, \(\reducedmodeldimension =390\)}
  \end{subfigure}
  \begin{subfigure}[t]{0.32\linewidth}\centering
  \includegraphics[width=\textwidth]{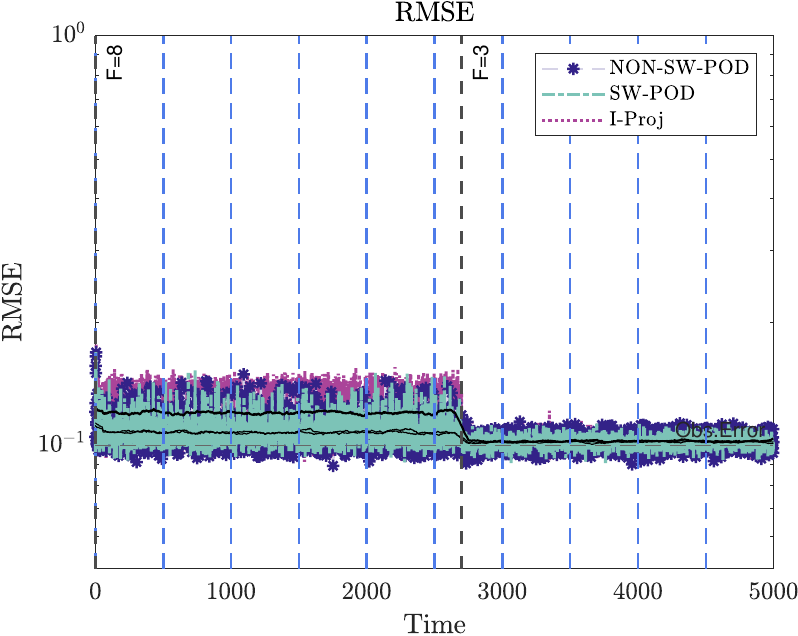}
  \caption{\gls{RMSE}, \(\reducedmodeldimension =400\)}
  \end{subfigure}
  \begin{subfigure}[t]{0.32\linewidth}\centering \includegraphics[width=\textwidth]{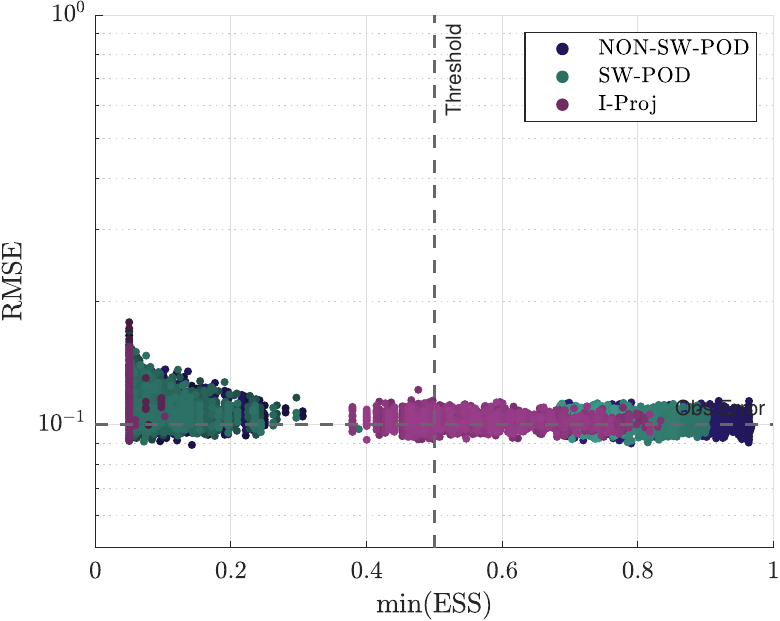}
  \caption{Trade-off between \gls{RMSE} and min(\gls{ESS}), \(\reducedmodeldimension =400\)}
  \end{subfigure}
  \begin{subfigure}[t]{0.32\linewidth}\centering
  \includegraphics[width=\textwidth]{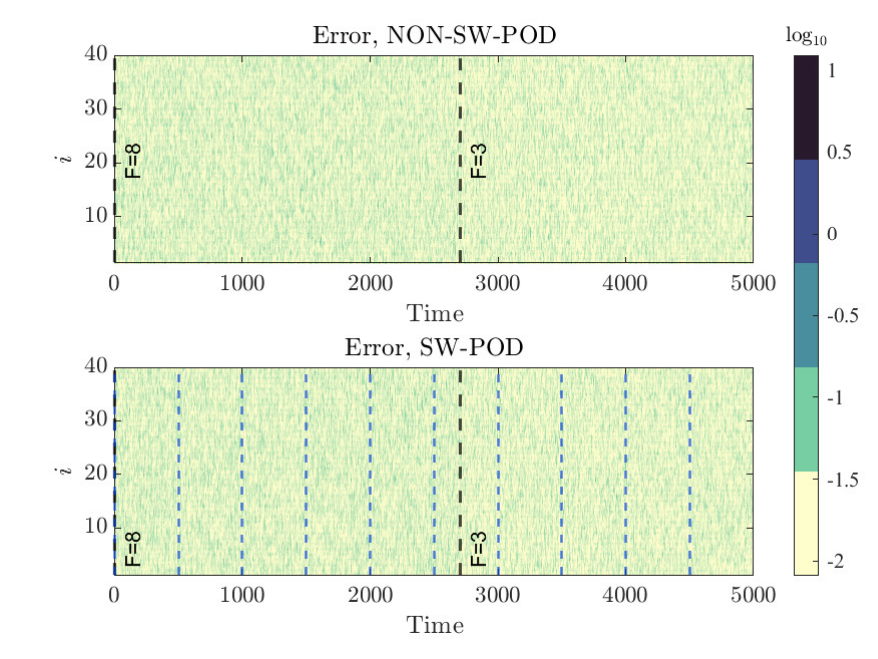}
  \caption{Error differences, \(\reducedmodeldimension =400\)}
  \end{subfigure}
   \caption{The effect of increasing the reduced physical model dimension to \(\reducedmodeldimension =390\) and \(\reducedmodeldimension =400\) (no reduction).
 The reduced data model \(\reduceddatadimension=5\).
  and the forcing parameter \(F\) is changing from \(8\rightarrow 3\) at \(\timeswitch=2700\) of final time \(\finaltime=5000\).}
  \label{fig:l96-exp4a}
\end{figure}

 We present experiments to show the robustness of mode selection using \gls{SWPOD}.
 We first experimented with  using the fixed dimensions of reduced models 
 with modes updated whenever the data window is moved.
 In general, the assimilation of the order reduction is challenging in the chaotic regime and using \gls{SWPOD} with a fixed dimension \gls{SWPOD} does not improve it.
We see only improvement using \gls{SWPOD} when we increase the size of the reduced physical model to very high dimensions \(\reducedmodeldimension=390\) or even \(\reducedmodeldimension=400\) resulting in a high error in the regular regime, see \cref{fig:l96-exp4a}.
This also demonstrates that when dynamics changes the regime but the order reduction is \emph{not} recomputed (as in \gls{NOSW}), the performance of the assimilation is heavily dependent on the duration of each regime.
Additionally, we found that shrinking the window size results in a faster adaptation of the estimate to the new dynamical regime, as long as the smaller window is efficacious at representing the dynamics, as is the case in the regular regime of \gls{L96} (\(F=3\)).

 Next, we changed the \gls{ROM} dimensions dynamically with every change in the data window by using a fixed fraction of the squared singular values and a sparse selection technique developed for the projected data model~(see \cref{sec:adaptive-selection-of-parameters} for the description).

The great advantage of using the \gls{SWPOD} is the fast adapting of the change of the forcing parameter \(F\).
 This allows for a change in rank of the projections to mimic the local Kaplan-Yorke dimension (see, e.g., \cite{QOKK2020}).
Overall, the offline fixed dimension \gls{SWPOD} is much more effective in a regular regime than the \gls{NOSW}.
Still, the assimilation with order reduction is challenging in the chaotic regime ($F=8$) with or without using the \gls{SWPOD}.

To overcome some of the inherent difficulties in having mode selection at fixed times, especially in the chaotic regime, the offline adaptivity technique (described in \cref{sec:offline-adapt}) determines the subspace dimension of \gls{SVD} condition to some tolerance value.
In earlier development of our techniques~\cite{MVV20,albarakati2021model} projected methods successfully lowered the dimension of the observation space.
We further augment the model selection for the observations by an online  sparse selection technique (described in \cref{sec:online-adapt}).
While still restricted to the larger sets of modes we have as candidates, the modes used for the projected data model are potentially sparsely selected.

In experiments (\(1-4\)), we present the result of the \gls{SWPOD} using offline and online adaptivity.
While the impact of the interaction between mode selection times and parameter switching is evident, the performance is still good in terms of both \gls{RMSE} and resampling.
In all experiments (\(1-4\)), good \gls{RMSE} has been achieved by tuning the adaptation parameters, with the added benefit of the adaptive online method (\(2-4\)), which results in the lowest resampling rate and the highest size reduction.

\subsection{Experiment 1: the effect of adaptive offline tolerance}\label{sec:offline-ex1}

\begin{figure}[ht!]
  \centering
  \begin{subfigure}[t]{0.45\linewidth}\centering
 \includegraphics[width=\textwidth]{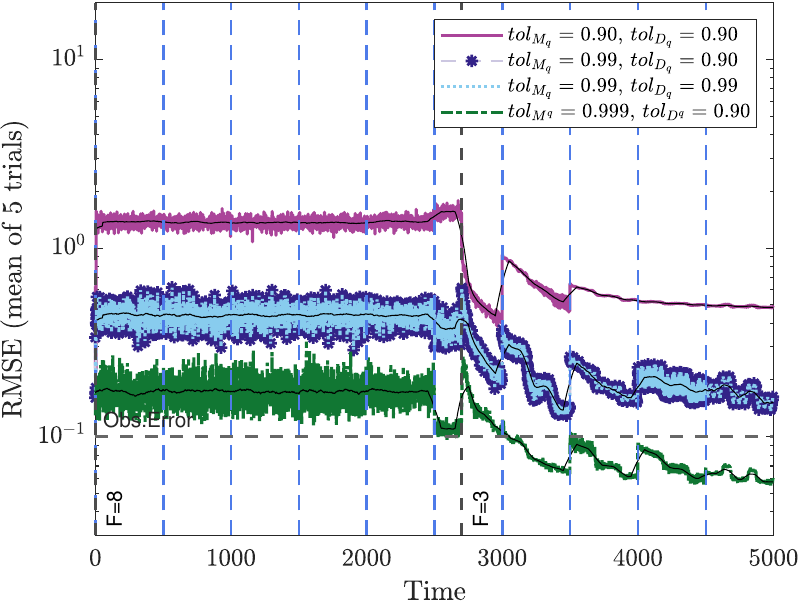}
  \caption{\gls{RMSE}}\label{fig:offline_a}
 \end{subfigure}
  \begin{subfigure}[t]{0.45\linewidth}\centering \includegraphics[width=\textwidth]{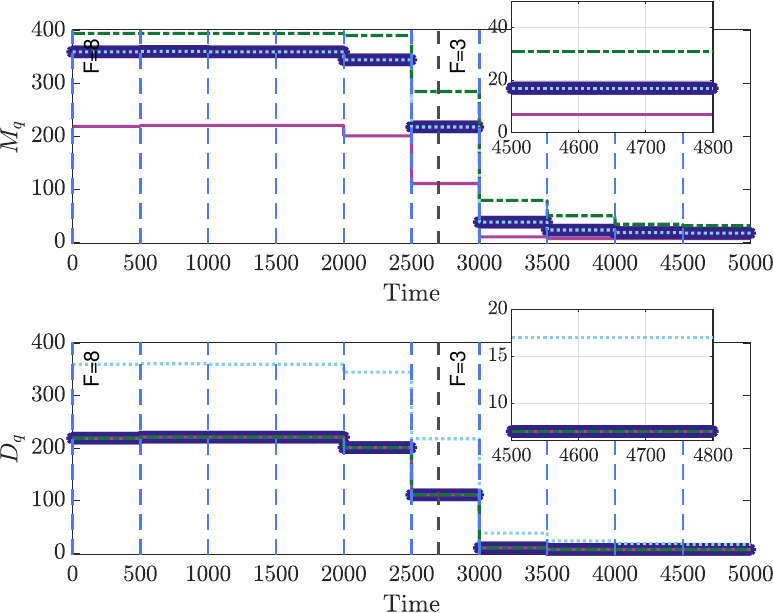}
  \caption{Sizes of the reduced model \(\reducedmodeldimension\) (left) and data \(\reduceddatadimension\) (right) dimensions }\label{fig:offline_b}
  \end{subfigure}
  \begin{subfigure}[t]{0.45\linewidth}\centering \includegraphics[width=\textwidth]{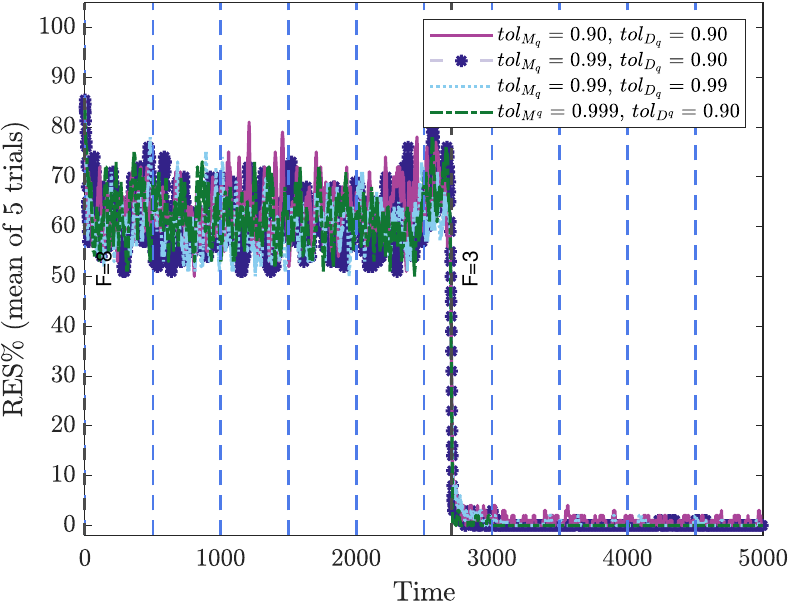}
  \caption{\gls{RESAMP}}\label{fig:offline_c}
  \end{subfigure}
\begin{subfigure}[t]{0.45\linewidth}\centering \includegraphics[width=\textwidth]{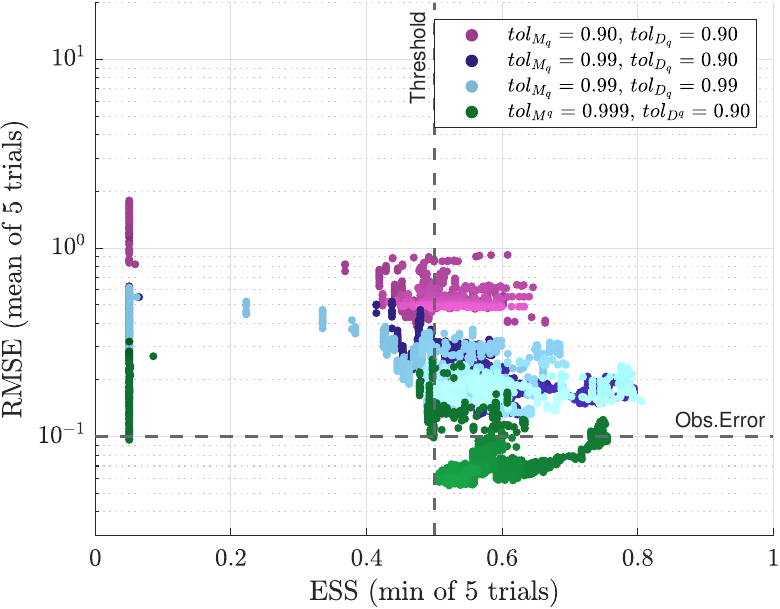}
  \caption{Trade-off between \gls{RMSE} and min(\gls{ESS})}
  \label{fig:RMSE_ESS_Adaptive}
  \end{subfigure}
 \caption{Experiment 1, offline-adaptive mode selection scheme (\cref{sec:offline-adapt}).
The reported value of the reduced model \(\reducedmodeldimension\) and data model \(\reduceddatadimension\) dimensions per window shift \(\windowshift=500\) are shown in (b).
The standard deviation of the observation noise (\(\dataerrorcovariance = 0.01\,\id\)) is given by the dashed line on \gls{RMSE} panels (a) and (d).
The threshold for the \gls{ESS} is set to be (\(\threshold=\frac{1}{2}\totalparticles\)) indicated by the vertical dash line on the second right panel. All numerical results in this experiment are obtained by averaging over five trials. See \cref{sec:offline-ex1} for further details.}
  \label{fig:l96-adaptive}
\end{figure}

\Cref{fig:l96-adaptive} shows the results of \gls{SWPOD} using the offline-adaptive mode selection scheme described in \cref{sec:offline-adapt}.
 
The offline adaptive mode selection method allows us to choose
the size of the reduced physical and data models for each window based on some tolerance value (i.e., the percentage of total information captured by the reduced space, \cref{eq:tol}). 
We consider three cases in which the data tolerance is fixed \(\tol_{\reduceddatadimension}=0.90\) and the physical model tolerances vary: \(\tol_{\reducedmodeldimension}=0.90\),  \(\tol_{\reducedmodeldimension}=0.99\) and \(\tol_{\reducedmodeldimension}=0.999\), and another case in which the data tolerance is increased to \(\tol_{\reduceddatadimension}=0.99\).
The color on \cref{fig:RMSE_ESS_Adaptive} changes from dark to light indicate the time changes, where we are looking for a low \gls{RMSE} and above threshold minimum \gls{ESS}.

Based on \gls{RMSE} in \cref{fig:offline_a}, the offline adaptive method converges more toward observation error as the tolerance of the offline model increases from \(\tol_{\reducedmodeldimension}=0.90\) to \(\tol_{\reducedmodeldimension}=0.999\) when the system is chaotic \(F=8\). In contracts, we do not observe a significant improvement in the \gls{RMSE} size when the data tolerance is increased to \(\tol_{\reduceddatadimension}=0.99\). In the overall offline adaptivity experiment, \(\tol_{\reduceddatadimension}=0.90\) appears to be sufficient where we do not benefit from larger values of high data tolerance but do benefit from larger values of model tolerance \(\tol_{\reducedmodeldimension}\).

For the size of the reduced dimensions model \(\reducedmodeldimension\) and the data \(\reduceddatadimension\) in \cref{fig:offline_b}, the offline adaptive mode selection achieves the minimum reduction in both the model and the data when the dynamic is more regular \(F=3\) with good \gls{RMSE}.In the chaotic regime \(F=8\), the sizes of both \(\reducedmodeldimension\) and \(\reduceddatadimension\) are still large (e.g., \(\reducedmodeldimension= 360,\,\reduceddatadimension=200\)).
That indicates the fixed mode selection needed \(\reducedmodeldimension \approx 400\) in the \(F=8\) region to get competitive \gls{RMSE} values as illustrated in \cref{fig:l96-exp4a}.
However, we do not see an improvement of the offline adaptivity mode selection scheme over the fixed mode selection scheme in terms of resembling in \cref{fig:offline_c} and \gls{ESS} \cref{fig:RMSE_ESS_Adaptive}, where the \gls{RESAMP} for both methods are still significant with lower than threshold \gls{ESS} in the chaotic regime.

To conclude, the offline adaptive mode selection scheme has the advantage of choosing the size of the reduced physical model dimension \(\reducedmodeldimension\) and data model dimension \(\reduceddatadimension\) dynamically, where we don't have to deal with overfitting and underfitting the reduced sizes. However, it's important to note that resampling and reduction in model and data dimensions remain significant in the chaotic regime.
\subsection{Experiment 2:  
the effect of adaptive online tolerance}\label{sec:online-ex1}

\begin{figure}[htb]
  \centering
  \begin{subfigure}[t]{0.45\linewidth}\centering \includegraphics[width=\textwidth, trim={0 0 0 15pt},clip
]{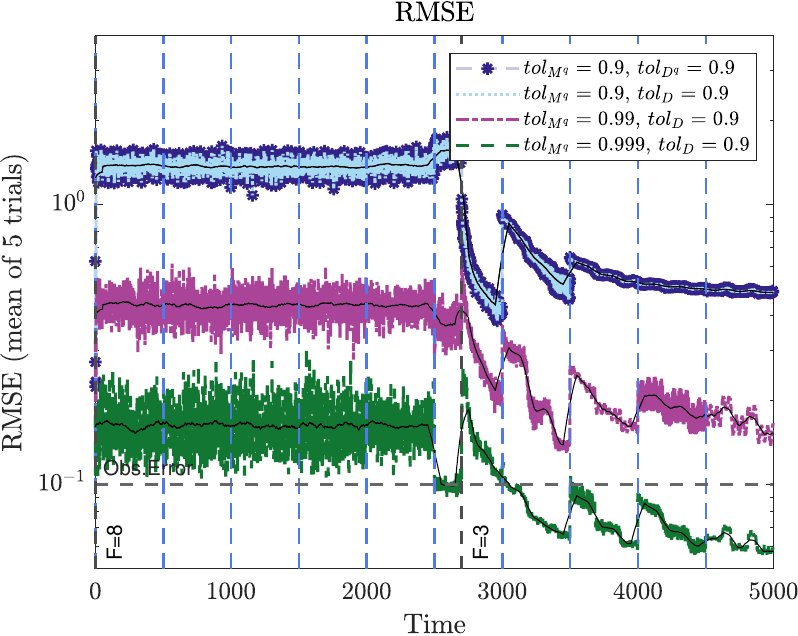}
  \caption{\gls{RMSE}}\label{fig:online-adapt_a}
\end{subfigure}
  \begin{subfigure}[t]{0.46\linewidth}\centering \includegraphics[width=\textwidth]{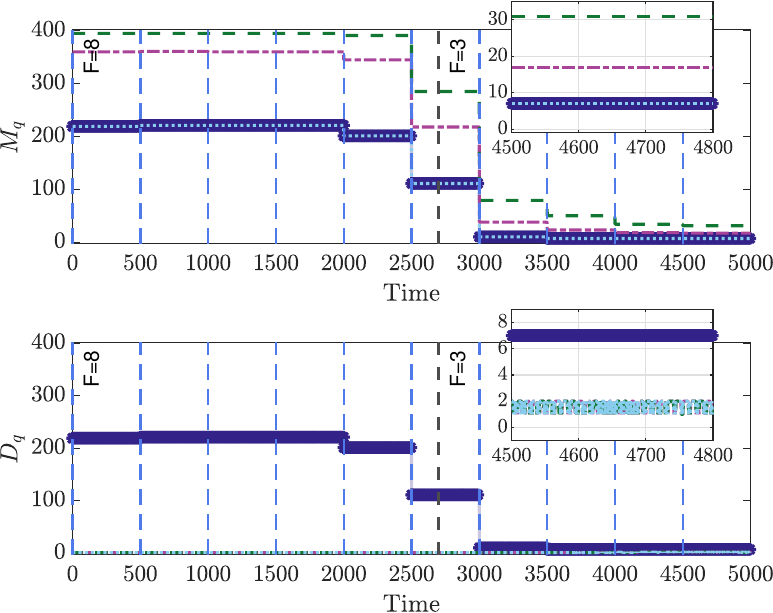}
  \caption{Sizes of the reduced model \(\reducedmodeldimension\) (top) and data \(\reduceddatadimension\) (bottom) dimensions averaging over five trials }\label{fig:online-adapt_b}
  \end{subfigure}
  \begin{subfigure}[t]{0.45\linewidth}\centering \includegraphics[width=\textwidth, trim={0 0 0 15pt},clip
]{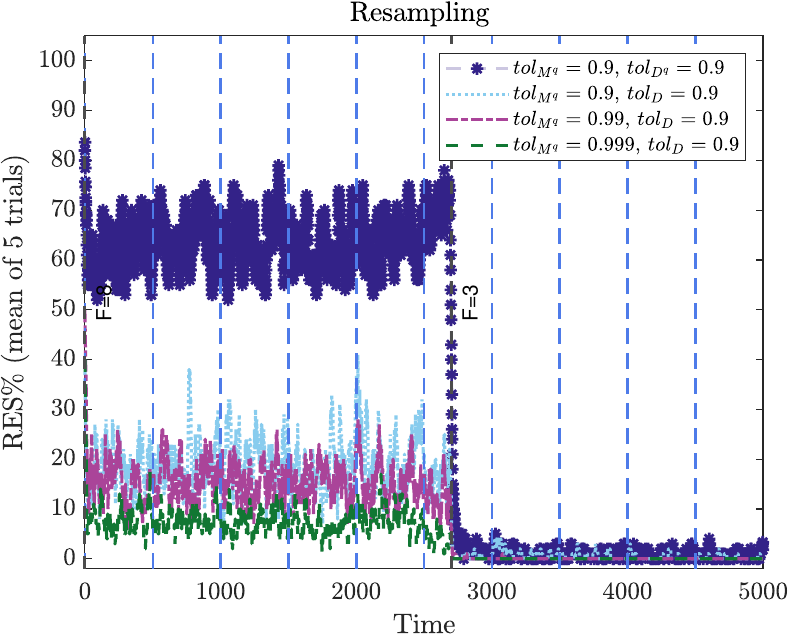}
  \caption{\gls{RESAMP}}\label{fig:online-adapt_c}
  \end{subfigure}
\begin{subfigure}[t]{0.45\linewidth}\centering \includegraphics[width=\textwidth]{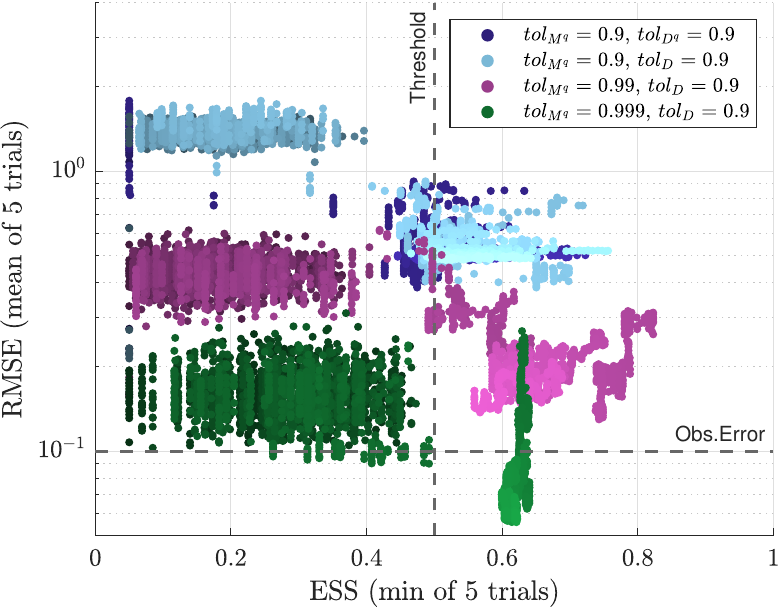}
 \caption{Trade-off between \gls{RMSE} and min(\gls{ESS})}\label{fig:online-adapt_d}
  \end{subfigure}
  \caption{Experiment 2, online adaptivity: the effect of adaptive online tolerance.
A comparison between \gls{SWPOD} the offline adaptivity with model's tolerance \(\tol_{\reducedmodeldimension}=0.9\) and data's tolerance \(\tol_{\reduceddatadimension}=0.9\), \cref{eq:tol}, and to the online adaptivity method with different offline tolerances for the model and tunable online tolerance \(\tol_{\datadimension}=0.9\) for the data, \cref{eq:lasso-regression}.
The standard deviation of the observation noise (\(\dataerrorcovariance = 0.01\,\id\)) is given by the dashed line in \gls{RMSE} panels.
 All numerical results in this experiment are obtained by averaging over five trials.
See \cref{sec:online-ex1} for further details.}
  \label{fig:online-adapt}
\end{figure}

\begin{figure}[htb]
\centering
\includegraphics[width=1.05\textwidth]{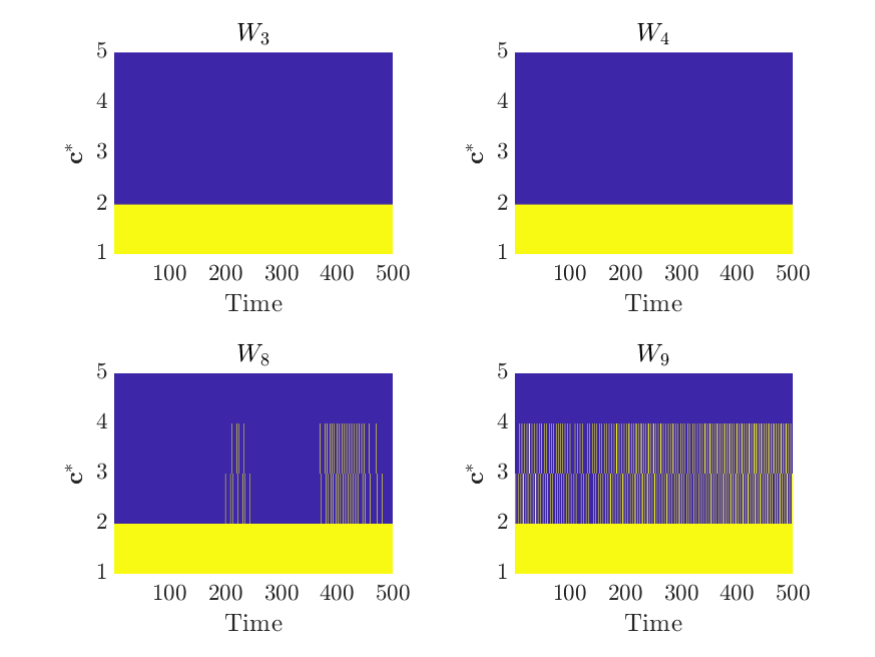}
\caption{The sparsity pattern for some selected time windows (\(W_i,\,i=3,4,8,9\)) with yellow color indicating the non-zero coefficients \(\vec{c}^\ast\) that correspond to the selected columns with value in \(\statereduction\) which are used to construct \(\datareduction\).
}\label{fig:sparsity}
\end{figure}

In this experiment, we compare \gls{SWPOD} using the online-adaptive mode selection scheme with physical model tolerances \(\tol_{\reducedmodeldimension}=\{0.9; 0.99; 0.999\}\), \cref{eq:tol} and tunable online tolerance \(\tol_{D}=0.9\), \cref{eq:lasso-regression}, to the offline mode selection scheme \cref{sec:offline-adapt}, with physical model tolerance \(\tol_{\reducedmodeldimension}=0.9\) and data model with tolerance \(\tol_{\reduceddatadimension}=0.9\).
\Cref{fig:sparsity} shows the indexes of non-zero coefficients \(\vec{c}^\ast\) that correspond to the columns in \(\statereduction\) that are used to construct \(\datareduction\) for some selected time windows \(W_i\), \(\{i=3,4,8,9\}\) of tunable online data tolerance \(\tol_{D}= 0.9\) and physical model tolerance \(\tol_{\reducedmodeldimension}=0.999 \).
We can see a pattern of the small size of the selected \(\statereduction\) modes indicated by the yellow color in \cref{fig:sparsity}.
Therefore, we expect a smaller overall dimension without the need to order the modes as the offline adaptivity, further reducing the reduced data dimension size.

The numerical results in \cref{fig:online-adapt} are obtained by averaging over five trials.
In terms of \gls{RMSE} in \cref{fig:online-adapt_a}, offline and online methods perform similarly, as you can see in the first two cases on the legend.
Despite both methods, offline \cref{fig:offline_a} and online \cref{fig:online-adapt_a}, show a convergent \gls{RMSE} to and below the observation error with increasing the offline tolerance, it is evident that the online modes selection scheme has superior performance when it comes to low resampling (\cref{fig:online-adapt_c}) and high \gls{ESS} above the threshold (\cref{fig:online-adapt_d}) when the system is chaotic (\(F=8\)).
The online adaptive mode selection scheme provides the lowest reduced data dimension (i.e.
\(\reduceddatadimension=1, 2\) uniformly in both regions \(F=8\) and \(F=3\)).
Similarly to the offline adaptivity experiment in \cref{sec:offline-ex1}, the online data adaptivity experiment does not benefit from larger values of high data tolerance \(\tol_{\reduceddatadimension}\) but from larger values of model tolerance \(\tol_{\reducedmodeldimension}\).

Overall, the online modes selection scheme with high offline model tolerance and online data tolerance (i.e., \(\tol_{\reducedmodeldimension}= 0.999\) \(\tol_{\reduceddatadimension}= 0.9\)) is the most effective method in comparison to all other methods that are offline fixed and adaptive dimension.
\subsection{Experiment 3: the effect of varying the observation time step of sparse observations}\label{sec:online-ex2}

\begin{figure}[t]
  \centering
  \begin{subfigure}[t]{0.45\linewidth}\centering \includegraphics[width=\textwidth]{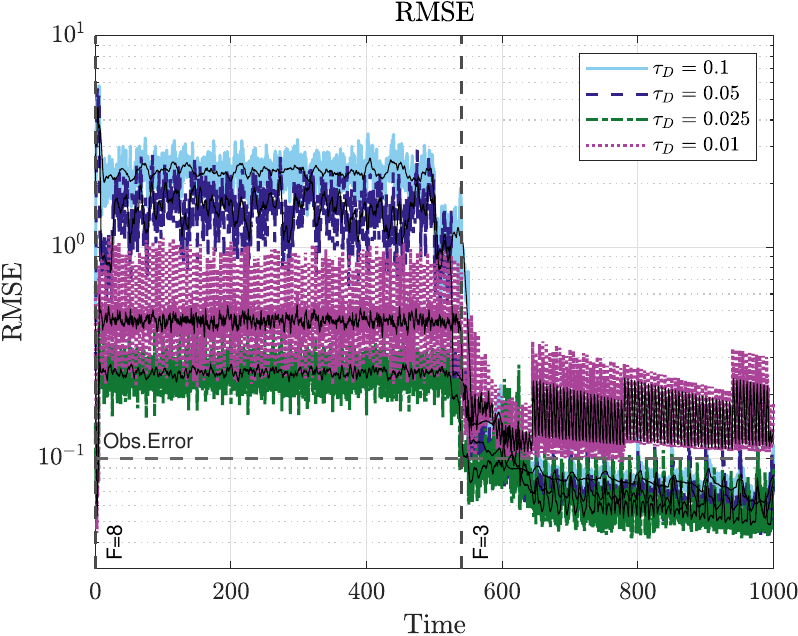}
  \caption{\gls{RMSE}, \(\dataoperator=\downsamp{2}\)}\label{fig:online-exp2b_a}
  \end{subfigure}
  \begin{subfigure}[t]{0.45\linewidth}\centering \includegraphics[width=\textwidth]{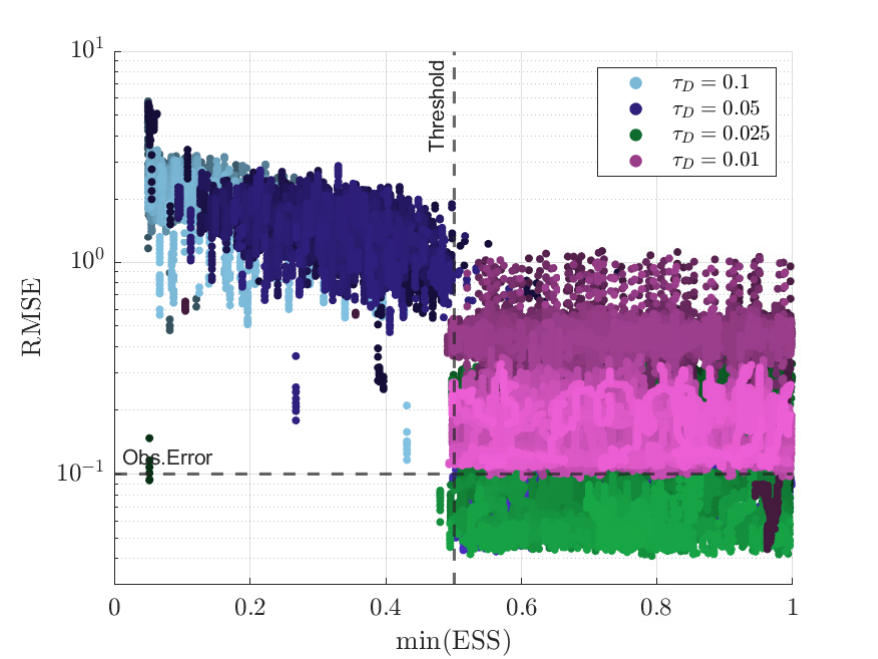}
  \caption{Trade-off between \gls{RMSE} and min(\gls{ESS}), \(\dataoperator=\downsamp{2}\)}\label{fig:online-exp2b_b}
  \end{subfigure}
  \begin{subfigure}[t]{0.45\linewidth}\centering \includegraphics[width=\textwidth]{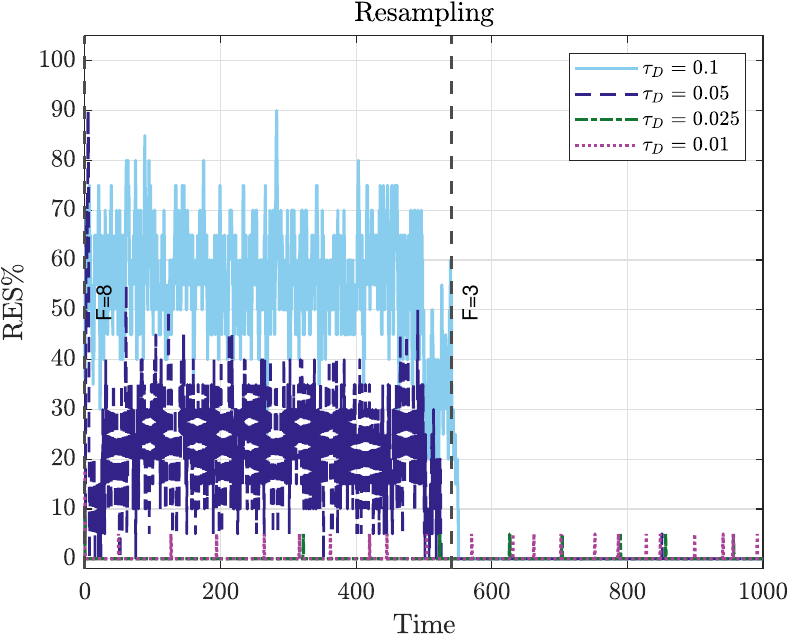}
  \caption{\gls{RESAMP},\, \(\dataoperator=\downsamp{2}\)}\label{fig:online-exp2b_c}
  \end{subfigure}
   \begin{subfigure}[t]{0.45\linewidth}\centering \includegraphics[width=\textwidth]{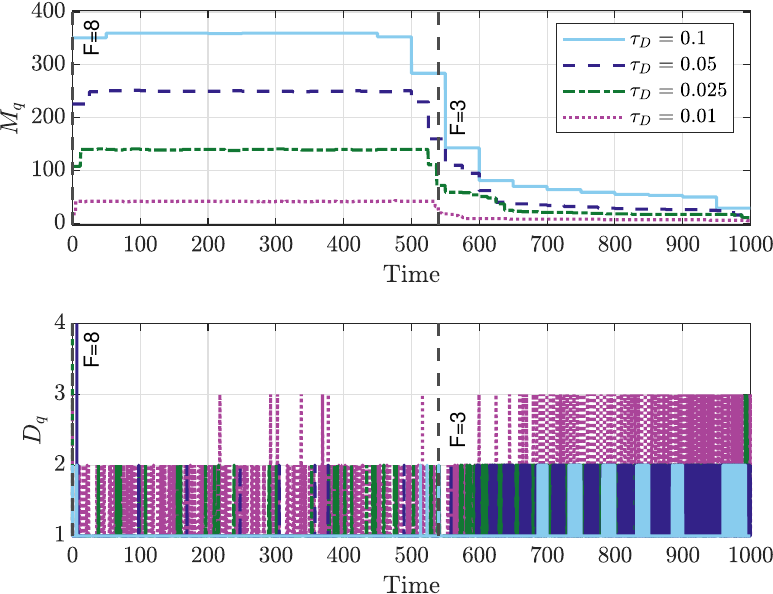}
  \caption{Sizes of the reduced model \(\reducedmodeldimension\) (top) and data \(\reduceddatadimension\) (bottom) dimensions, \(\dataoperator=\downsamp{2}\)}\label{fig:online-exp2b_d}
  \end{subfigure}
  \caption{Experiment 3, online adaptivity: the effect of sparse observation, \(\dataoperator=\downsamp{2}\), and varying the observation time step \(\timestepobs\) where \(\timestepobs =0.1; 0.05; 0.025; 0.01\) of online adaptivity scheme.
The online-adaptive mode selection scheme with offline physical model tolerances \(\tol_{\reducedmodeldimension}=0.999\), \cref{eq:tol} and tunable online tolerance \(\tol_{D}=0.9\), \cref{eq:lasso-regression}.
See \cref{sec:online-ex2} for further details.}
  \label{fig:online-exp2b}
\end{figure}

In this experiment \cref{fig:online-exp2b} shows the effect of varying the observation time step (\(\timestepobs =0.1; 0.05; 0.025; 0.01\)) of sparse observation (\(\dataoperator=\downsamp{2}\)) where \(\downsamp{2}\) is canonical projections onto every other state variable using the online adaptivity scheme.
The online-adaptive mode selection scheme with offline physical model tolerances \(\tol_{\reducedmodeldimension}=0.999\), \cref{eq:tol} and tunable online tolerance \(\tol_{D}=0.9\), \cref{eq:lasso-regression}.

The \gls{RMSE} in \cref{fig:online-exp2b_a} is converging to and below the observation error (\(\sqrt{\dataerrorcovariance}\)) in both regime (i.e., chaotic (\(F=8\)) and regular (\(F=3\))) of a sparse observation (\(\dataoperator=\downsamp{2}\)) with a smaller time step (i.e., \(\timestepobs=0.025\)).
The \gls{ESS} with smaller time steps (i.e., \(\timestepobs=0.025\)  and \(\timestepobs=0.01\)) of sparse observation (\(\dataoperator=\downsamp{2}\)) is staying above than the threshold as in \cref{fig:online-exp2b_b} which indicates less resampling as shown in \cref{fig:online-exp2b_c}.
The sizes of the reduced model dimension \(\reducedmodeldimension\), which are calculated adaptively offline, decrease as the observation time step \(\timestepobs\) gets smaller, as shown in the top panel of \cref{fig:online-exp2b_d}.
Whereas the sizes of the reduced data dimension \(\reduceddatadimension\) are calculated adaptively online using lasso regression \cref{eq:lasso-regression} (see \cref{sec:online-adapt} for the description).
As a result, the sizes of the reduced data dimension \(\reduceddatadimension\) are very small and stable no matter how small the observation time step \(\timestepobs\) as you can see in the lower panel of \cref{fig:online-exp2b_d}.

\begin{figure}[htb]
  \centering
  \begin{subfigure}[t]{0.32\linewidth}\centering \includegraphics[width=\textwidth]{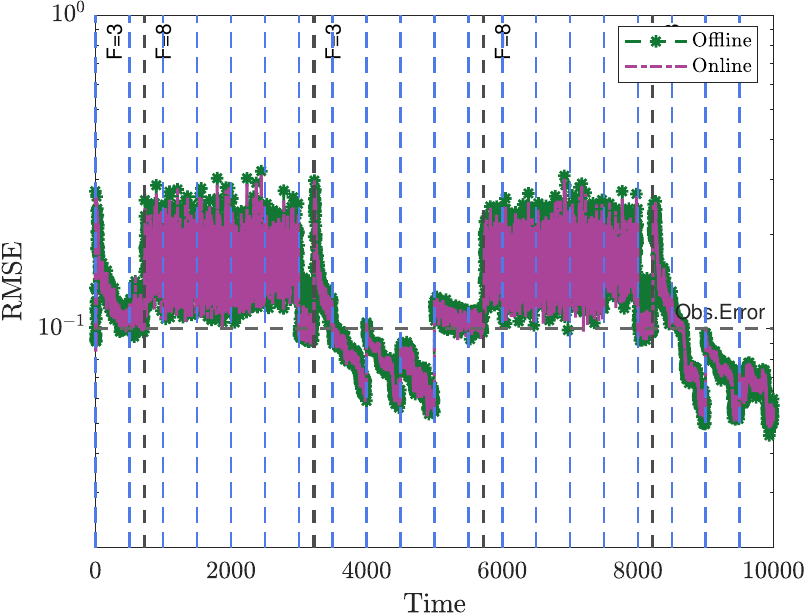}
  \caption{\gls{RMSE}}\label{fig:online-exp3_a}
  \end{subfigure}
    \begin{subfigure}[t]{0.32\linewidth}\centering \includegraphics[width=\textwidth]{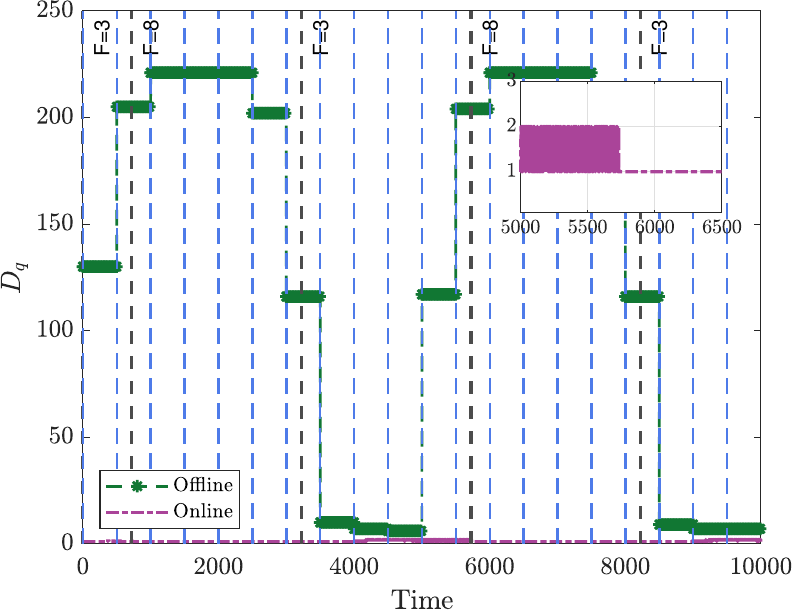}
  \caption{Sizes of the reduced model data \(\reduceddatadimension\) dimensions }\label{fig:online-exp3_b}
  \end{subfigure}
  \begin{subfigure}[t]{0.32\linewidth}\centering \includegraphics[width=\textwidth]{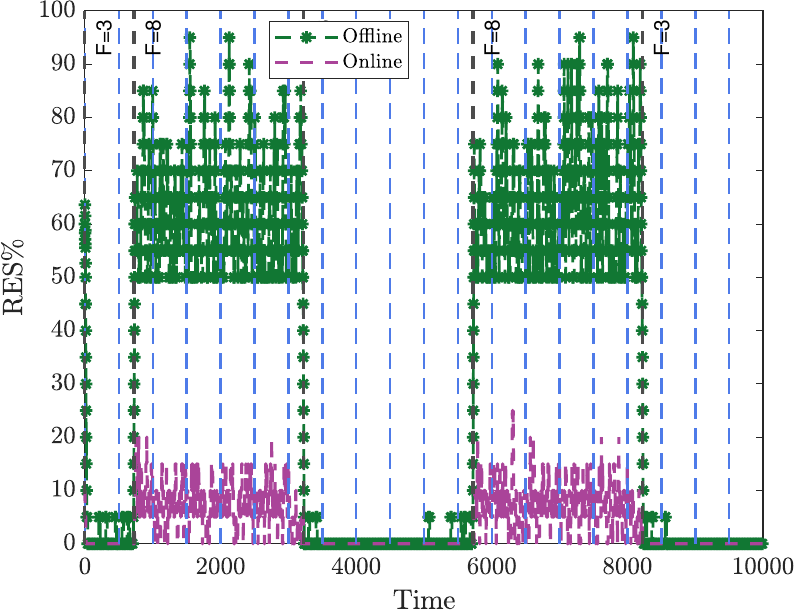}
  \caption{\gls{RESAMP}}\label{fig:online-exp3_c}
  \end{subfigure}
  \caption{Experiment 4, online adaptivity: the effect of time-dependent forcing parameter.
The \gls{L96} model dimension is \(\modeldimension=400\), final time \(\finaltime=10000\), the forcing parameter jumps from \(F = 8\rightarrow 3\) at \(\timeswitch=3224, 8224\), and from \(F= 3\rightarrow 8\) at \(\timeswitch=724, 5724\).
  The standard deviation of the observation noise (\(\dataerrorcovariance = 0.01\,\id\)) is given by the horizontal dash line in \gls{RMSE} panels.
See \cref{sec:online-ex3}  for further details.}
  \label{fig:online-exp3}
\end{figure}
\subsection{Experiment 4: the effect of time-dependent forcing parameter of online adaptivity}\label{sec:online-ex3}

In this experiment, we compare the offline adaptivity (\cref{sec:offline-adapt}) and online adaptivity (\cref{sec:online-adapt}) to see how the time-dependent forcing parameter \(F\) affects the simulation.
The forcing term \(F\) in \eqref{eq:l96} determines whether the evolution will be qualitatively regular or chaotic.
This experiment uses an offline adaptive mode selection scheme based on offline tolerances for model \(\tol_{\reducedmodeldimension}=0.999\), and data \(\tol_{\reduceddatadimension}=0.90\), \cref{eq:tol}.
The online-adaptive mode selection scheme based on offline physical model tolerances \(\tol_{\reducedmodeldimension}=0.999\), \cref{eq:tol} and tunable online tolerance \(\tol_{D}=0.9\), \cref{eq:lasso-regression}.

The effects of changing dynamics from regular to chaotic and vice versa at different time switches are shown in \cref{fig:online-exp3}.
Both adaptive online and offline methods provide a convergent \gls{RMSE} below the observation error whether the system is regular or chaotic as in \cref{fig:online-exp3_a}.
The adaptive online method wins by offering the most significant reduction in data dimension in \cref{fig:online-exp3_b} with a maximum reduced data dimension \(\reduceddatadimension=2\) and with the lowest percentage of resampling in \cref{fig:online-exp3_c} where the highest total number of times resampling has occurred for offline.

Overall, the adaptive online method provides the best results with the lowest \gls{RMSE}, \gls{RESAMP} and reduction of data dimension \(\reduceddatadimension\).

\section{Discussion and Conclusion}
In this paper, we show the efficiency of using \gls{SWPOD} with the developed \gls{ProjOPPF} in \cite{albarakati2021model}, \cite{MVV20}.
Generally, the \gls{ProjOPPF} is developed to perform well if either the physical model or the observational data have a smaller effective dimension or both.
A low resampling percentage and \gls{RMSE} can be achieved using lower-dimensional projected models if the adequate dimensions are sufficiently small.
The \gls{SWPOD} with fixed and adaptive mode selection methods show promising results with lower \gls{RMSE} and \gls{RESAMP}, smaller error differences, and higher \gls{ESS} than \gls{NOSW}.
In addition, \gls{SWPOD} reacts faster to the time-varying forcing parameter \(F= 8\rightarrow 3\) of \gls{L96}, where we saw a quick drop in \gls{RMSE} in all experiments in the same window as \(F\) changes.
The \gls{SWPOD} with the offline adaptive mode selection method described in~\cref{sec:offline-adapt}, shows successful results with a high model tolerance value \(\tol_{\reducedmodeldimension}=0.999\).

The \gls{SWPOD} with an adaptive online method described in~\cref{sec:online-adapt} performs the best out of all other methods (i.e., fixed offline and adaptive offline) in terms of low \gls{RMSE}, less resampling \gls{RESAMP} and high \gls{ESS}.
It also offers the most reduction in the data dimension, as we saw in experiments (6-8).
The techniques developed are effective through a combination of very low adaptive data dimension and model dimension that are optimized dynamically dependent on the underlying complexity of the behavior of the physical model.

As a future work, we are exploring using a sliding window with \glsreset{DMD}\gls{DMD} applied to the two-layer Lorenz 96 coupled model with changing coupling parameters.
Another avenue is the development of adaptive in-time techniques to determine when to update modes based on monitoring the representation error in projecting onto the current set of modes.
\appendix
\printglossaries
\bibliographystyle{elsarticle-num}
\bibliography{references}

\end{document}